%% file: paper.tex
\documentclass[submission, Phys]{SciPost}

\pdfoutput=1
\input{incl_settings.tex}   

\input{incl_shortcuts.tex}          
\graphicspath{{./figs/}}                

\begin{document}

\begin{center}{\Large \textbf{
Lorentz-Equivariance without Limitations
}}\end{center}

\begin{center}
Luigi Favaro\textsuperscript{1},
Gerrit Gerhartz\textsuperscript{2},
Fred A. Hamprecht\textsuperscript{3},
Peter Lippmann\textsuperscript{3},\\
Sebastian Pitz\textsuperscript{2},
Tilman Plehn\textsuperscript{2,3},
Huilin Qu\textsuperscript{4},
and Jonas Spinner\textsuperscript{2}
\end{center}

\begin{center}
{\bf 1} CP3, Universit\'e catholique de Louvain, Louvain-la-Neuve, Belgium \\
{\bf 2} Institut für Theoretische Physik, Universität Heidelberg, Germany \\
{\bf 3} Interdisciplinary Center for Scientific Computing (IWR), Universität Heidelberg, Germany
{\bf 4} CERN, EP Department, Geneva, Switzerland
\end{center}

\begin{center}
\today
\end{center}

\section*{Abstract}
         {Lorentz Local Canonicalization (LLoCa) ensures exact Lorentz-equivariance for arbitrary neural networks with minimal computational overhead. For the LHC, it equivariantly predicts local reference frames for each particle and propagates any-order tensorial information between them. We apply it to graph networks and transformers. We showcase its cutting-edge performance on amplitude regression, end-to-end event generation, and jet tagging. For jet tagging, we introduce a large top tagging dataset to benchmark LLoCa versions of a range of established benchmark architectures and highlight the importance of symmetry breaking.}

\vspace{2pt}
\noindent\rule{\textwidth}{1pt}
\tableofcontents\thispagestyle{fancy}
\noindent\rule{\textwidth}{1pt}

\section{Introduction}
\label{sec:intro}

Modern machine learning (ML) is increasingly shaping the present and future of the physics program at the LHC. In the search of new physics and precise measurements, ML techniques are being deployed in the simulation pipeline from phase space sampling~\cite{Heimel:2022wyj,Janssen:2023lgz} and parton distribution functions~\cite{NNPDF:2021njg} to fast detector simulation~\cite{ATLAS:2021pzo}.
In view of the HL-LHC they show great promise for jet tagging~\cite{Kasieczka:2019dbj}, fast detector simulation~\cite{Krause:2024avx}, anomaly detection~\cite{CMS:2024nsz, ATLAS:2025kuz, Hallin:2021wme, Das:2023bcj, Favaro:2023xdl}, unfolding~\cite{Huetsch:2024quz, Butter:2024vbx, Andreassen:2019cjw, Favaro:2025psi}, and event generation~\cite{Butter:2023fov,Butter:2024zbd}. In particular, graph neural networks (GNN) and transformers provide powerful architectures which respect the structure of particle data~\cite{Qu:2019gqs,Qu:2022mxj,Vent:2025ddm}. 

To fully exploit the structure of LHC data and achieve maximum performance, we need to embed the relevant fundamental symmetries in ML networks. To this end, Lorentz-equivariance or covariance is of prime interest. Several Lorentz-equivariant architectures have boosted the performance of jet taggers~\cite{Butter:2017cot, Gong:2022lye,Bogatskiy:2023nnw, ruhe2023clifford, Spinner:2024hjm, Brehmer:2024yqw}. More recently, they have been extended to amplitude surrogates~\cite{Breso:2024jlt,Brehmer:2024yqw}, event generation~\cite{Brehmer:2024yqw} and BSM searches~\cite{Cruz:2024grk,Qureshi:2025dma}. However, equivariant approaches are often task specific, not easily transferable, and for problems that are symmetric only under a subgroup of the Lorentz group, symmetry breaking becomes necessary. For these reasons, it remains unclear how the performance of Lorentz-equivariant networks with symmetry breaking compares to networks that are only equivariant under a subgroup of the Lorentz group~\cite{Bhardwaj:2024djv,Maitre:2024hzp,Chatterjee:2024pbp,Nabat:2024nce}.

Lorentz Local Canonicalization (LLoCa) provides a way to upgrade any network to an exact Lorentz-equivariant version, as proposed for a ML audience in Ref.~\cite{Spinner:2024hjm}. It builds upon the concept of local canonicalization~\cite{lippmann2025beyond} and extends it to Lorentz symmetry. In the LLoCa framework, we predict local reference frames for each input particle and exploit Lorentz-invariance in the local frames to define latent particle features of any geometric representation order. When applying LLoCa to message passing networks, like GNNs or transformers, the frame-to-frame transformations include full geometric message passing information with arbitrary representations of the Lorentz group. 
Because the equivariant prediction of local frames requires negligible resources, we achieve a flexible and computation-efficient implementation of Lorentz-equivariance without the need for specific building blocks. As the task-specific backbone is untouched, we can provide a fair assessment of the performance gain from Lorentz-equivariance. 

We test LLoCa on a comprehensive set of LHC applications from the regression of squared matrix-element scattering amplitudes to the generation of high-multiplicity final states, and jet tagging. Going beyond Ref.~\cite{Spinner:2025prg}, we present a generative network based on LLoCa and benchmark LLoCa taggers on a new, large-scale top tagging dataset. We discuss the results with a special focus on Lorentz symmetry breaking and the impact of internal representations, supported by several performance studies. In particular, LLoCa allows us to directly compare Lorentz-equivariant networks using symmetry breaking with $\mathrm{SO}(2)$-equivariant networks when only a residual $\mathrm{SO}(2)$-symmetry is realized in the data. The corresponding data augmentation can easily be implemented in LLoCa as the sampling of random global frames assigned to the entire point cloud. We perform an in-depth comparison with non-equivariant neural networks trained with data augmentation. 
 
We start by describing the LLoCa formalism in Sec.~\ref{sec:formalism}, followed by a detailed description of the construction of local reference frames and Lorentz symmetry breaking. As the first application, we discuss surrogate amplitudes for $Z\;+$ gluons in Sec.~\ref{sec:amplitudes}. In Sec.~\ref{sec:generation} we introduce a generative network for high-multiplicity LHC events. Finally, we perform extensive studies for jet tagging in Sec.~\ref{sec:tagging}. In particular, we train LLoCa versions of established high-performance taggers and compare the performance and the training times on the top tagging and the JetClass datasets~\cite{Qu:2022mxj}.

\section{Lorentz Local Canonicalization}
\label{sec:formalism}
According to special relativity, the laws of physics are invariant under the choice of inertial reference frame when related by Lorentz transformations.
In this section, we discuss Lorentz Local Canonicalization (LLoCa)~\cite{Spinner:2025prg}, 
a framework to construct Lorentz-equivariant adaptations of arbitrary network architectures using a learnable reference frame for each particle and tensorial message passing between frames. In this paper we provide an expanded and physics-motivated introduction with a particular focus on Lorentz symmetry breaking.

\subsection{Local reference frames}
\label{sec:local_frames}

Starting with a Lorentz vector $x$ in a global reference frame, we can define a Lorentz transformation $L \in \mathbb{R}^{4\times 4}$ to map $x \to x_L=Lx$, such that $x_L$ is invariant under a general Lorentz transformations $\Lambda$
\begin{align}
    x_L \overset{\Lambda}{\longrightarrow} x'_L = x_L.
    \label{eq:xL}
\end{align}
The Lorentz transformations $L$ and $\Lambda$ satisfy the defining property $L^TgL = g, \Lambda^Tg\Lambda = g$ with the Minkowski metric $g=\mathrm{diag}(+,-,-,-)$. 
Inserting $x_L=Lx$ and using $x\to x' = \Lambda x$ gives us the transformation rule of the local frame $L$ under $\Lambda$
\begin{align}
    L \overset{\Lambda}{\to} L' = L \Lambda^{-1}\quad\text{such that}\quad 
    x_L \overset{\Lambda}{\to} x'_L = L' x' = L \Lambda^{-1} \Lambda x = L x = x_L.
    \label{eq:local_frame}
\end{align}

One way to construct local reference frames is from a set of four orthonormal unit vectors ${e_a}^\mu$ with ${e_a}^\mu {e_b}^\nu g_{\mu\nu} = g_{ab}$ ($a,b=0,1,2,3$). We define a local frame as a matrix with the covectors ${e^a}_{\mu} = g^{ab}{e_b}^\nu g_{\nu\mu}$ as rows,
\begin{align}
    L = \begin{pmatrix}
    \rule[0.5ex]{0.5cm}{0.55pt} \ g{e_0^Tg} \ \rule[0.5ex]{0.5cm}{0.55pt}\\
    \rule[0.5ex]{0.5cm}{0.55pt} \ g{e_1^Tg} \ \rule[0.5ex]{0.5cm}{0.55pt}\\
    \rule[0.5ex]{0.5cm}{0.55pt} \ g{e_2^Tg} \ \rule[0.5ex]{0.5cm}{0.55pt}\\
    \rule[0.5ex]{0.5cm}{0.55pt} \ g{e_3^Tg} \ \rule[0.5ex]{0.5cm}{0.55pt}\\
    \end{pmatrix}
    \label{eq:L_from_vectors}
\end{align}
The definition of local frames based on unit vectors allows us to write them as matrices,
\begin{align}
  {L^a}_\mu = {e^a}_\mu
  \qquad \text{with} \qquad a,\mu = 0,1,2,3 \; ,
  \label{eq:L_compact}
\end{align}
and easily derive their transformation rule. Above, we distinguish between the Lorentz greek index $\mu$ and the Roman index $a$, where the latter indicates an invariant index that does not mix under Lorentz transformations. Using this notation, Eq.~\eqref{eq:local_frame} becomes
\begin{align} \label{eq:local_frame_indices}
    {L^a}_\mu \overset{\Lambda}{\to} {L'^a}_\mu = {L^a}_\nu {(\Lambda^{-1})^\nu}_\mu \; .
\end{align}
Local frames constructed as in Eq.~\eqref{eq:L_compact} satisfy the required transformation behaviour of Eq.~\eqref{eq:local_frame_indices} since 
\begin{align}
  L{^a}_\mu = e{^a}_\mu \quad  \overset{\Lambda}{\to}  \quad  
    L'{^a}_\mu = e'{^a}_\mu = e{^a}_\nu(\Lambda^{-1}){^\nu}_\mu = {L^a}_\nu {(\Lambda^{-1})^\nu}_\mu \;,
\end{align}
as by definition of the transformation rule of covectors, $x'_\mu = x_\nu {(\Lambda^{-1})^\nu}_\mu$. Moreover, since the local and the global reference frames are both flat spaces and we consider orthonormal vectors ${e_a}^\mu$, the local frames $L$ act as Lorentz transformations.
By introducing invariant indices $a,b,c\dots$ we can read off the transformation behavior of Lorentz tensors in the LLoCa formalism from their index structure, for example
\begin{alignat}{7}
  x^\mu &\overset{\Lambda}{\to} {\Lambda^\mu}_\nu x^\nu
  &\qqquad
  x_\mu &\overset{\Lambda}{\to} x_\nu {(\Lambda^{-1})^\nu}_\mu \notag 
  &\qqquad
  f^{\mu\nu}_\lambda &\overset{\Lambda}\to {\Lambda^\mu}_\sigma {\Lambda^\nu}_\kappa f^{\sigma\kappa}_\gamma {(\Lambda^{-1})^\gamma}_\lambda\\
  x_L^a &\overset{\Lambda}{\to} x_L^a
  &\qqquad
  (x_L)_a &\overset{\Lambda}{\to} (x_L)_a
  &\qqquad
  (f_L)^{ab}_c &\overset{\Lambda}{\to} (f_L)^{ab}_c \; .
\end{alignat}
Note that general relativity uses the same ``vierbein'' formalism to construct locally flat coordinate systems as $g_{\mu\nu} = g_{ab} {L^a}_\mu {L^b}_\nu$, where $g_{ab}$ is the standard Minkowski metric while $g_{\mu\nu}$ is the metric tensor in curved spacetime~\cite{Weinberg:1972kfs}.

\subsection{Lorentz-equivariance from local canonicalization}
\label{sec:local_canonicalization}

To illustrate the principle of local canonicalization in the context of Lorentz-equivariant architectures, we first focus on the case of a vector input $x^\mu$.
Transforming this input into the local frame yields the invariant object
\begin{align}
  \text{input:} \qquad x_L^a = {L^a}_\mu x^\mu \; .
\end{align}
As an invariant, we can embed $x_L^a$ into a latent space and process it with unconstrained operations, including standard linear layers, dropout, layer normalization, attention, etc.
To output another vector, we extract a four-dimensional object $y_L^a$ from the backbone architecture and we transform it back into the global frame,
\begin{align}
  \text{output:} \qquad y^\mu = {(L^{-1})^\mu}_a y_L^a \; .
\end{align}
Using Eq.~\eqref{eq:local_frame} and the fact that the network output in the local frame is invariant, $y'^a_L = y^a_L$, we find that the output vector $y^\mu$ satisfies the correct transformation behavior
\begin{align}
  y^\mu \overset{\Lambda}{\to} y'^\mu
  &= {(L'^{-1})^\mu}_a y'^a_L  \notag \\
  &= {\Lambda^\mu}_\nu {(L^{-1})^\nu}_a y^a_L = {\Lambda^\mu}_\nu y^\nu.
    \label{eq:vector_equivariance}
\end{align}
This example shows that, while other Lorentz-equivariant architectures rely on specialized layers, LLoCa can incorporate any backbone architecture and still produce an exact Lorentz-equivariant output. Next, we show that the output feature can be of arbitrary representation.

To generalize to arbitrary representations for the input $x$ and output $y$ only the tensor transformation behavior has to be modified. To characterize how elements from the symmetry group act on node features or, more generally, elements from a vector space, we employ the notion of a group representation. For a general group $G$ and vector space $V$, a representation is a group homomorphism $\rho: G \to \mathrm{GL}(V)$, i.e.~a mapping that respects the group structure via $\rho (a b) = \rho (a)\rho (b)$ for all $a,b \in G$. We denote the action of a general representations of the Lorentz group on $x$ as
\begin{align}
  x' = \rho_x (L) x \; .
\end{align}
For Lorentz tensors, the representation is expressed through their index structure, e.g.~for a third-order tensor $x^{\mu\nu\lambda}$,
\begin{align}
x'^{\mu\nu\lambda} = \rho_x(x)^{\mu\nu\lambda}_{\sigma\kappa\eta} x^{\sigma\kappa\eta} \quad \mathrm{with}\quad \rho_x(x)^{\mu\nu\lambda}_{\sigma\kappa\eta} = {\Lambda^\mu}_\sigma {\Lambda^\nu}_\kappa {\Lambda^\lambda}_\eta.
\end{align}
For any group representations, the invariance of the input features in the local frame $x_L$ and the equivariance of the output features $y$ follow directly as generalizations of Eq.~\eqref{eq:local_frame} and Eq.~\eqref{eq:vector_equivariance}:
\begin{align}
    x_L = \rho (L) x &\overset{\Lambda}{\to} \rho (L \Lambda^{-1}) \rho (\Lambda ) x = \rho (L \Lambda^{-1} \Lambda )x = \rho (L) x = x_L \notag \\
    y = \rho (L^{-1}) y_L &\overset{\Lambda}{\to} \rho (\Lambda L^{-1}) y_L = \rho (\Lambda ) \rho(L^{-1})y_L = \rho (\Lambda )y \; .
    \label{eq:lloca_equivariance}
\end{align}
In contrast to Lorentz-equivariant architectures relying on specialized layers like 
LorentzNet~\cite{Gong:2022lye}, PELICAN~\cite{Bogatskiy:2023nnw} and L-GATr~\cite{Spinner:2024hjm}, LLoCa supports arbitrary higher-order input, output, and latent representations.

\subsection{Message Passing between local reference frames}
\label{sec:message_passing}

LHC data is typically a point cloud of particles, so one would naturally assign distinct local frames to each particle. Architectures like graph networks and transformers process particles separately and exchange information between them through message passing~\cite{bronstein2021geometric} of the following form
\begin{align}
    f^\text{updated}_i = \psi \left( f_i, \smboxop_{j=1}^N \phi (m_j)\right) \; .
    \label{eq:message-passing}
\end{align}
We write $\psi,\phi$ for unconstrained neural networks and $\smboxop$ for a permutation-invariant aggregation function such as the component-wise mean or maximum. The message $m_j = m_j(f_j)$ is constructed from the latent features $f_j$.
These messages have their own representation $\rho_m$ which is treated as a hyperparameter of the architecture. Its choice has no effect on the general formalism outlined above, but it defines the tensorial nature of the messages exchanged between different local frames, i.e.~how many Lorentz tensors of a specific order can be communicated.

\begin{figure}[t]
 \centering
   \includegraphics[width=0.6\linewidth]{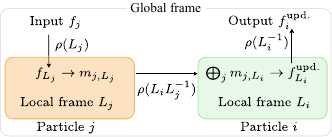}
   \caption{Tensorial message passing in LLoCa. All symbols are described in the text. This figure is also included in Ref.~\cite{Spinner:2025prg}.}
   \label{fig:message-passing}
\end{figure}
We consider the general example of sending a message from particle $j$ to particle $i$, with the corresponding local frames $L_j$ and $L_i$. The tensorial message passing in LLoCa is illustrated in Figure~\ref{fig:message-passing}. By construction, $m_{j,L_j}$ is invariant in the sender frame $L_j$. We gather the corresponding $m_{j,L_i}$ in the frame $L_i$, so the two messages are related by the frame transformation matrix $L_i L_j^{-1}$
\begin{align}
    m_{j,L_i} = \rho_m (L_i L_j^{-1}) m_{j,L_j}\;.
    \label{eq:tensorial_message}
\end{align}
For only scalar representations in $\rho_m$ this reduces to $m_{j,L_i}=m_{j,L_j}$, reverting back to Eq.~\eqref{eq:message-passing}. For our applications, we find that the network performance reduces drastically in this scenario because the network cannot meaningfully communicate tensorial information between frames.

Instead, we typically use a mix of scalar and vector representations in $\rho_m$ although our implementation also supports arbitrary higher-order representations, see Table~\ref{tab:ampxl_abl_messages}. Regardless of the representation $\rho_m$, the received message $m_{j,L_i}$ is invariant since the frame transformation matrix $L_i L_j^{-1}$ is invariant
\begin{align}
  {(L_i)^a}_\mu {(L_j^{-1})^\mu}_b
  \overset{\Lambda}{\to} {(L'_i)^a}_\mu {(L'^{-1}_j)^\mu}_b
  &= {(L_i)^a}_\nu {(\Lambda^{-1})^\nu}_\mu {\Lambda^\mu}_\lambda {(L_j^{-1})^\lambda}_b \notag \\
  &= {(L_i)^a}_\mu {(L_j^{-1})^\mu}_b\;.
\end{align}
Including the frame-to-frame transformation of the messages in Eq.~\eqref{eq:tensorial_message}, we generalize Eq.~\eqref{eq:message-passing} for message passing architectures in the LLoCa framework to
\begin{align}
  f^\text{updated}_{i,L_i}
  &= \psi \left( f_{i,L_i}, \smboxop_{j=1}^N \phi \left(m_{j,L_i} \right)\right) \notag \\
  &= \psi \left( f_{i,L_i}, \smboxop_{j=1}^N \phi \left(\rho_m (L_i L_j^{-1}) m_{j,L_j} \right)\right)\;.
    \label{eq:tensorial_message_passing}
\end{align}
The scaled dot-product attention used in transformers~\cite{vaswani2017attention} emerges as a special case of Eq.~\eqref{eq:tensorial_message}. Adopting the recipe of Eq.~\eqref{eq:tensorial_message_passing}, we obtain the scaled dot-product attention in the LLoCa framework
\begin{align}
  f^\text{updated}_{i,L_i}
  &= \sum_{j=1}^N \text{softmax} \left(\frac{1}{\sqrt{d}} \langle q_{i,L_i}, k_{j,L_i}\rangle\right) v_{j,L_i} \notag \\
    &= \sum_{j=1}^N \text{softmax} \left(\frac{1}{\sqrt{d}} \langle q_{i,L_i}, \rho_k (L_i L_j^{-1}) k_{j,L_j}\rangle \right) \rho_v (L_i L_j^{-1}) v_{j,L_j}\;. 
\end{align}
The queries $q_{L_i}$, keys $k_{L_j}$ and values $v_{L_j}$ are $d$-dimensional objects computed from the latent features in the respective local frame. The Minkowski product $\langle \cdot ,\cdot \rangle$ contracts all open indices with the metric tensor, e.g.~$\langle q, k\rangle = q^{\mu\nu} g_{\mu\lambda} g_{\nu\sigma} k^{\lambda\sigma}$ for second-order tensors. Finally, using the invariance of the Minkowski product under Lorentz transformations we find
\begin{align}
  \Langle q_{i,L_i}, \rho_k (L_iL_j^{-1}) k_{j,L_j}\Rangle
  &= \Langle \rho_k (L_i^{-1} q_{i,L_i}, \rho_k (L_i^{-1}) \rho_k (L_i L_j^{-1}) k_{j,L_j}\Rangle \notag \\
  &= \Langle \rho_k (L_i^{-1}) q_{i,L_i}, \rho_k (L_j^{-1}) k_{j,L_j}\Rangle \; .
\end{align}
In other words, instead of transforming the sender features to the receiving frame to compute the attention scores, we can first transform all the features into the global frame, compute the scaled dot-product attention, and then transform the result back into the receiving frame. This allows us to use optimized attention implementations~\cite{dao2022flashattention}.

\subsection{Constructing local frames}
\label{sec:construct_frames}

Next, we show how to construct local reference frames $L$ that satisfy
\begin{align}
  L^TgL=g
  \qquad \text{and} \qquad
  L' = L \Lambda^{-1} \; ,
\end{align}
from Sec.~\ref{sec:local_frames}. We first construct a set of Lorentz vectors $v_k^\mu$ from the input particles using a Lorentz-equivariant architecture, and then orthonormalize them to obtain a valid Lorentz transformation.

Consider a system of $N$ particles, each one described by its four-momentum $p_i$ and additional scalar information $s_i$, e.g.~the particle type. We use a simple Lorentz-equivariant architecture to construct a set of three vectors $v_{i,k}^\mu$ for each particle, 
\begin{align}
  v_{i,k}^\mu = \sum_{j=1}^N \text{softmax} \left( \varphi_k (s_i, s_j, \langle p_i,p_j\rangle )\right) \frac{p_i^\mu+p_j^\mu}{\|p_i+p_j\| + \epsilon}
  \qqquad  (k=0,1,2) \; ,
    \label{eq:construct_v}
\end{align}
with $\|p\| = \sqrt{p_\mu p^\mu}$. The local frames network (Frames-Net) $\varphi_k$ is an MLP operating on scalar features. Eq.~\eqref{eq:construct_v} ensures that the vectors $v_{i,k}^\mu$, being a linear combination of Lorentz vectors with scalar coefficients, are predicted in a Lorentz-equivariant way~\cite{villar2021scalars}. The $\text{softmax}$ operation ensures that all coefficients are positive. Together with four-momenta $p_i$ with positive norm, this ensures positively normed vectors $v_{i,k}^\mu$.

The design of Eq.~\eqref{eq:construct_v} involves only a single message-passing operation and is therefore limited to pairwise information exchange. More expressive Frames-Net architectures based on multi-layer graph networks or transformers overcome this limitation; we discuss two such options in Section~\ref{sec:framesnet-alternatives}. However, in our studies, we found that the simple MLP approach of Eq.~\eqref{eq:construct_v} consistently achieves good performance.

Positive norms, and hence timelike vectors, are essential for the numerical stability of our applications. Simpler choices such as $\exp$ and $\text{softplus}$ yield a comparable but slightly worse performance.
Similarly, rescaling the particle four-momenta as $(p_i+p_j)/(\|p_i+p_j\| + \epsilon)$ slightly improves numerical stability compared to a simple factor of $p_j$. In general, we design Eq.~\eqref{eq:construct_v} to control the prediction of boosted vectors $v_{i,k}^\mu$ at initialization, as they lead to boosted local frames $L$, which can affect numerical stability. 

Given a set of three vectors $v_k$, we construct the local frame $L$ for each particle independently. We use a polar decomposition theorem to factorize it into a rotation and a pure boost,
\begin{align}
    L = R B.
    \label{eq:polar_decomposition}
\end{align}
The rotation and boost are constructed separately. First, the boost is constructed from the vector $v_0$ as
\begin{align}
    B = \begin{pmatrix}
        \gamma & -\gamma\vec\beta^T \\
        -\gamma\vec\beta & I_3 + (\gamma -1)\dfrac{\vec\beta\vec\beta^T}{\vec\beta^2}
    \end{pmatrix} \qquad\text{with} \qquad
    \beta^i &= \frac{v_0^i}{\| v_0\|} \notag \\
    \gamma &= \frac{v_0^0}{\|v_0\|} = (1 - \vec\beta^2)^{-1/2}.
\label{eq:boost_parametrization}
\end{align}
This requires $\|v_0\|>0$, guaranteed by the $\text{softmax}$ in Eq.~\eqref{eq:construct_v}. We then extract the boosted vectors $w_k = Bv_k$ for $k=1,2$. Afterwards, the Gram-Schmidt algorithm is used to extract a set of orthonormal three-vectors $\vec e_1, \vec e_2, \vec e_3$ from the input three-vectors $\vec w_k = (\vec w_1, \vec w_2)$
\begin{align}
    \operatorname{norm}(\vec w) = \frac{\vec w}{\|\vec w\| + \epsilon} \qquad 
    \vec e_1 &= \operatorname{norm}\left(\vec w_1 \right),\notag \\
    \vec e_2 &= \operatorname{norm}\left(\vec w_2 - \vec e_1 (\vec w_2 \cdot \vec e_1)\right), \notag\\
    \vec e_3 &= \vec e_1 \times \vec e_2.
    \label{eq:gram-schmidt}
\end{align}
The resulting unit vectors are combined to the rotation matrix
\begin{align}
    R = \begin{pmatrix}
        1 & \vec 0^T \\ \vec 0 & \tilde R
    \end{pmatrix}
    \qquad \text{with} \qquad 
    \tilde R = (\vec e_1, \vec e_2, \vec e_3)^T.
\end{align}
This procedure is equivalent to a Gram-Schmidt algorithm in Minkowski space~\cite{Spinner:2025prg}. However, we find this approach to be numerically more stable.

\subsection{Advanced Frames-Net architectures}
\label{sec:framesnet-alternatives}

To construct frames with the correct transformation behavior in Eq.~\eqref{eq:local_frame}, LLoCa requires a Lorentz-equivariant network to predict vectors $v_{i,k}^\mu$.
Our default Frames-Net, described in Sec.~\ref{sec:construct_frames}, employs a simple MLP $\varphi_k$ acting on Lorentz-invariant features.
To stabilize the training, it is essential to constrain this network to predict only timelike vectors. The timelike nature of the input particle four-momenta leads to the specific configuration of Eq.~\eqref{eq:construct_v} to ensure that the output vectors $v_{i,k}^\mu$ are timelike,
\begin{align}
  v_{i,k}^\mu = \sum_{j=1}^N \text{softmax} \left( \varphi_k (s_i, s_j, \langle p_i,p_j\rangle )\right) \frac{p_i^\mu+p_j^\mu}{\|p_i+p_j\| + \epsilon}
  \qqquad  (k=0,1,2) \; .
\end{align}
Although we find this straightforward approach sufficient for all applications considered in this work, the timelike constraint prohibits the out-of-the-box usage of already existing Lorentz-equivariant architectures.
In the following, we present two Frames-Nets based on the PELICAN architecture and L-GATr.
These two advanced Frames-Nets address two potential limitations of our default design:
\begin{itemize}
    \item The edge convolution in Eq.~\eqref{eq:construct_v} allows only a single message-passing step within the Frames-Net architecture. This limits its expressivity and may hinder performance in tasks with complex interaction patterns.
    \item For events with a large number of particles, fully connected graph networks such as the edge convolution in Eq.~\eqref{eq:construct_v} can incur higher memory costs than transformer-based architectures. Consequently, if the backbone network is a transformer, the Frames-Net may contribute noticeably to the overall memory usage, even though its parameter count and runtime overhead remain small.
\end{itemize}
The PELICAN architecture can address the first issue, while the L-GATr architecture offers a remedy for both.

\subsubsection*{PELICAN Frames-Net}

The PELICAN architecture naturally generalizes an MLP operating on edge features by introducing message-passing steps between edges. While the simple construction of Eq.~\eqref{eq:construct_v} includes only a single message-passing step through the edge convolution, a PELICAN-based Frames-Net allows for multiple such steps,
\begin{align}
  v_{i,k}^\mu = \sum_{j=1}^N \text{softmax} \left( \mathrm{PELICAN}^{n=2}_k \left(\{s_l\}_{l=1}^N, \{\langle p_n,p_m\rangle\}_{m=1,n=1}^N \right)_{ij} \right) \frac{p_i^\mu+p_j^\mu}{\|p_i+p_j\| + \epsilon} \; .
    \label{eq:construct_v2}
\end{align}
Here, $\mathrm{PELICAN}_k^n$ denotes a PELICAN network as described in App.~\ref{app:pelican}, which embeds node-level features $s_l$ into edge-level features and processes them jointly with the Lorentz-invariant edge features $\langle p_n, p_m \rangle$. In contrast to Eq.~\eqref{eq:construct_v}, the coefficients $c_{ij}$ entering the softmax now depend on the entire event, i.e.~the full set of scalar features $\{s_l\}_{l=1}^N$ and all pairwise inner products $\{\langle p_n,p_m\rangle\}_{m=1,n=1}^N$. The PELICAN network consists of a series of $\mathrm{Eq}_{2\to 2}$ blocks, each providing a complete set of permutation-equivariant aggregation operations that enable information exchange across edge features. The resulting edge representations are then used as Lorentz-invariant weights to construct the vectors $v_{i,k}^\mu$ in Eq.~\eqref{eq:construct_v2}.
In our implementation, we employ the PELICAN-lite variant described in App.~\ref{app:pelican}, with 1 block with 32 hidden channels, and 16 hidden channels for the aggregation maps, totalling 12k parameters.

\subsubsection*{L-GATr Frames-Net}

To overcome the memory limitations of graph-based networks in the Frames-Net, we employ a transformer architecture. In particular, we must avoid learnable operations on edges such as those in $\varphi_k(s_i, s_j, \langle p_i, p_j \rangle)$ and $\mathrm{PELICAN}_k^{n=2}(s_i, \langle p_i, p_j \rangle)$. A natural approach is to reinterpret Eq.~\eqref{eq:construct_v} as a generalized attention mechanism, where the value is a fixed, non-learnable matrix $p_i^\mu + p_j^\mu$ in $(i, j)$. This leads to
\begin{align}
  q_{i,k}, k_{i,k} &= \mathrm{LayerNorm}\circ \text{L-GATr}_k(\{s_l\}_{l=1}^N,\{p_n\}_{n=1}^N)_i\; ,  & \notag\\
  v_{i,k}^\mu &= \sum_{j=1}^N \text{softmax} \left( \sum_{c=1}^{n_c} \frac{\langle q_{i,k,c}, k_{j,k,c}\rangle}{\sqrt{16 n_c}} \right) (p_i^\mu + p_j^\mu) \; .
    \label{eq:construct_v3}
\end{align}
Again, the coefficients $c_{ij}$ entering the softmax depend on the full event, through the scalar features $\{s_l\}_{l=1}^N$ and the four-momenta $\{p_n\}_{n=1}^N$ provided to the L-GATr network.
The numbers of scalar and multivector channels in $q_{i,k}$ and $k_{i,k}$ are treated as hyperparameters, which we set equal to the latent dimensionality of $\mathrm{L\text{-}GATr}$. This L-GATr-based Frames-Net thus avoids learnable edge operations, substantially reducing memory usage for large point clouds.
We find that applying the multivector $\mathrm{LayerNorm}$~\cite{Brehmer:2024yqw} further improves training stability. We omit the rescaling by $(\|p_i + p_j\| + \epsilon)$, as it tends to destabilize training. As in Eqs.~\eqref{eq:construct_v} and \eqref{eq:construct_v2}, using $p_i^\mu + p_j^\mu$ instead of $p_j^\mu$ improves stability by preventing large boosts in the resulting frames.
Finally, note that directly extracting the vectors $v_{i,k}^\mu$ from the vector-grade output of L-GATr is not a viable alternative, as this approach does not guarantee that the resulting vectors are timelike.
In our implementation, we use 1 L-GATr block with 2 multivector and 24 scalar channels, totalling 12k parameters.

\medskip
We tested the PELICAN and L-GATr Frames-Net architectures for the regression and classification tasks and found that they achieve similar or slightly worse performance compared to the MLP Frames-Net baseline. We attribute this finding to suboptimal tuning of design choices in Eqs.~\eqref{eq:construct_v2},~\eqref{eq:construct_v3}.

\subsection{Lorentz symmetry breaking}
\label{sec:symmetry_breaking}

In LHC applications, symmetry is often realized with respect to only a subgroup $\mathcal{R}$ of the full Lorentz group.
For instance, detector effects, jet clustering algorithms, and the beam axis direction reduce the symmetry group to the $\mathrm{SO}(2)_\text{beam}$ group of rotations around the beam axis.

A fully Lorentz-equivariant network would disregard these effects and, therefore, symmetry breaking has to be included to achieve optimal performance.
The flexibility of the LLoCa framework allows for a fair comparison between different Lorentz-symmetry breaking approaches, which is difficult to achieve with previous Lorentz-equivariant architectures.

\subsubsection*{Lorentz symmetry breaking through architecture}

First, we change the architecture such that Eq.~\eqref{eq:lloca_equivariance} only holds for $\Lambda\in\mathcal{R}$. While this is challenging to achieve in Lorentz-equivariant architectures with specialized layers~\cite{Gong:2022lye, Bogatskiy:2023nnw, Spinner:2024hjm}, LLoCa allows to reduce the symmetry group by fixing a subset of reference directions $v_k$ in the construction of local frames in Eq.~\eqref{eq:construct_v}
\begin{itemize}
    \item $\mathcal{R}=\mathrm{SO}(3)$: $v_0 = (1,0,0,0)$;
    \item $\mathcal{R}=\mathrm{SO}^+(1,1)_\text{beam}\times \mathrm{SO}(2)_\text{beam}$: ${v_0}^1 = {v_0}^2 = 0, v_1 = (0, 1, 0, 0)$;
    \item $\mathcal{R}=\mathrm{SO}(2)_\text{beam}$: $v_0=(1,0,0,0)$, $v_1=(0,0,0,1)$;
    \item $\mathcal{R}=1$ (non-equivariant): $v_0=(1,0,0,0)$, $v_1=(0,0,0,1)$, $v_2 = (0, 1,0,0)$ or $L = 1$.
\end{itemize}
Intuitively, this means that all the frames agree on a set of global directions, e.g.~the time direction $(1,0,0,0)$ or the beam direction $(0,0,0,1)$. Note that a non-zero time component in the beam direction for the $\mathrm{SO}(2)_\text{beam}$-equivariant network would be automatically removed in the Gram-Schmidt algorithm. The group $\mathrm{SO}^+(1,1)_\text{beam}\times \mathrm{SO}(2)_\text{beam}$ of boosts along the beam axis and rotations around the beam axis is isomorph to the group of translations in $(\eta, \phi)$, with the azimuthal angle $\phi$and the rapidity $\eta$. Fixing all possible global directions corresponds to a non-equivariant architecture.

\subsubsection*{Lorentz symmetry breaking through input}

Alternatively, we break Lorentz-equivariance at the input level. This can be done for any architecture and is typically used in Lorentz-equivariant architectures relying on specialized layers~\cite{Gong:2022lye, Bogatskiy:2023nnw, Spinner:2024hjm}. Here, we add extra inputs which are only invariant under $\mathcal{R}$. 
For instance, the beam or time direction can be added as reference vectors $r$, with a similar assignment to the residual symmetry group $\mathcal{R}$ as in the listing above. They transform non-trivially under general Lorentz transformations, but fixing them as $(1,0,0,0)$ and $(1,0,0,1)$ breaks this transformation behaviour. However, residual $\mathrm{SO}(2)_\text{beam}$ transformations leave the time and beam directions invariant. Concretely, we use the following reference vectors to break Lorentz equivariance down to equivariance under $\mathcal{R}$:
\begin{itemize}
    \item $\mathcal{R}=\mathrm{SO}(3)$: $r_0 = (1,0,0,0)$;
    \item $\mathcal{R}=\mathrm{SO}^+(1,1)_\text{beam}\times \mathrm{SO}(2)_\text{beam}$: $r_0=(1,0,0,1)$, $r_2=(1,0,0,-1)$;
    \item $\mathcal{R}=\mathrm{SO}(2)_\text{beam}$: $r_0=(1,0,0,0)$, $r_1=(1,0,0,1)$, $r_2=(1,0,0,-1)$;
    \item $\mathcal{R}=1$ (non-equivariant): $r_0=(1,0,0,0)$, $r_1=(1,1,0,0)$, $r_2=(1,0,1,0)$, $r_3=(1,0,0,1)$.
\end{itemize}
Alternatively, one can include auxiliary inputs like the energy $E$ or transverse momentum $p_T$, which behave like scalars under the residual $\mathcal{R}$ group and therefore preserve equivariance. For jet tagging, we include the following scalar features:
\begin{itemize}
    \item $\mathcal{R}=\mathrm{SO}(3)$: $E$;
    \item $\mathcal{R}=\mathrm{SO}^+(1,1)_\text{beam}\times \mathrm{SO}(2)_\text{beam}$: $p_T$, $\Delta R$, $\Delta\phi$, $\Delta\eta$;
    \item $\mathcal{R}=\mathrm{SO}(2)_\text{beam}$: $E, p_T, \Delta R$, $\Delta\phi$, $\Delta\eta$;
    \item $\mathcal{R}=1$ (non-equivariant): $E, p_T, \Delta R$, $\Delta\phi$, $\Delta\eta$.
\end{itemize}
These two kinds of symmetry-breaking inputs are redundant. For example, for $\mathcal{R}=\mathrm{SO}(3)$ the energy $E$ can be recovered from the reference vector $r_0=(1,0,0,0)$ via $E = \langle r_0, p\rangle$. Nevertheless, using both types of inputs can improve performance in practice. For LLoCa networks, it is sufficient to add either type of symmetry-breaking input to the Frames-Net in Eq.~\eqref{eq:construct_v}.

In contrast to the hard breaking at the architecture level, symmetry breaking at the input level can be dynamically tuned during training by ignoring the extra inputs. This makes it powerful in practice, because the neural network can learn to restore full Lorentz-equivariance in phase space regions where it is useful.
The event generation, in Sec.~\ref{sec:generation}, and the jet tagging application, in Sec.~\ref{sec:tagging}, require symmetry breaking and will be used to compare the two approaches.

\section{Amplitude regression with LLoCa}
\label{sec:amplitudes}

Our first application of Lorentz Local Canonicalization (LLoCa) is amplitude regression.\footnote{The results in this section were first presented in Ref.~\cite{Spinner:2025prg}.}
Scattering amplitudes can be expressed analytically as functions of the particle four-momenta. Their numerical evaluation becomes expensive at higher loop orders or with  many external particles. In such cases, amplitude surrogates can significantly speed up event generation~\cite{Aylett-Bullock:2021hmo,Maitre:2021uaa,Badger:2022hwf,Maitre:2023dqz,Brehmer:2024yqw,Breso:2024jlt,Bahl:2024gyt}. The challenge is to provide networks that are sufficiently expressive to learn the amplitude from a limited number of training points such that the prediction error is much smaller than the other uncertainties arising from theory calculations.

We benchmark the LLoCa networks for the partonic processes
\begin{align}
    q\bar q \to Z + n g \qquad \text{with} \qquad  n = 1... 4 \; .
\end{align}
We use MadGraph~\cite{madgraph} to generate $10^7$ training events for $n=1,2,3$, and $10^8$ training events for $n=4$, a considerably larger statistics than in Ref.~\cite{Brehmer:2024yqw}.
For each multiplicity, we also generate $5\times 10^5$ events for evaluation and $10^5$ events for validation. We begin with a standard MadGraph run to generate unweighted phase space points, and then apply the standalone module to evaluate the corresponding squared amplitudes. To avoid divergences, we apply global cuts for all final-state particles.
\begin{align}
    p_T > 20\;\text{GeV}\qquad \text{and}\qquad \Delta R>0.4\;.
\end{align}
We use the MSE of the standardized logarithmic amplitudes as a loss function, 
\begin{align}
    \loss = \left\langle \left(\mathcal{A}_\text{surrogate} - \mathcal{A}_\text{true}\right)^2\right\rangle
    \qquad \text{with} \qquad 
    \mathcal{A} = \frac{\log A - \overline{\log A}}{\sigma_{\log A}}
\end{align} \; .
For evaluation, we use the same MSE as performance metric for simple comparison across datasets. The input four-momenta are rescaled by the standard deviation of the entire dataset.

To achieve sufficient precision, the symmetries present in the amplitudes should be represented in the network~\cite{Brehmer:2024yqw,Breso:2024jlt}. As baselines, we consider two architectures that enforce Lorentz-invariant outputs:
\begin{itemize}
    \item MLP-I~\cite{Breso:2024jlt}, a simple MLP with Lorentz-invariant inputs $p_ip_j$;
    \item L-GATr~\cite{Brehmer:2024yqw}, a transformer operating on Lorentz group representations.
\end{itemize}
While L-GATr also enforces the permutation symmetry of identical particles present in the amplitude, MLP-I has to learn this property from data. To demonstrate the flexibility of LLoCa, we further consider two permutation-equivariant architectures, 
\begin{itemize}
    \item a standard transformer~\cite{Brehmer:2024yqw};
    \item a message passing graph network (GNN) which uses Lorentz-invariant edge features $\langle p_i,p_j\rangle$.
\end{itemize}
LLoCa allows us to turn both architectures Lorentz-equivariant. We use the same Frames-Net architecture to predict the local frames for the LLoCa-Transformer and LLoCa-GNN, using only 20k learnable parameters. Details about the implementation and training can be found in Appendix~\ref{app:training}.

\subsection*{Results}
\sisetup{detect-weight=true, input-symbols = {( )}}
\begin{figure}[t!]
    \begin{minipage}[t!]{0.46\linewidth}\centering
    \includegraphics[width=\linewidth]{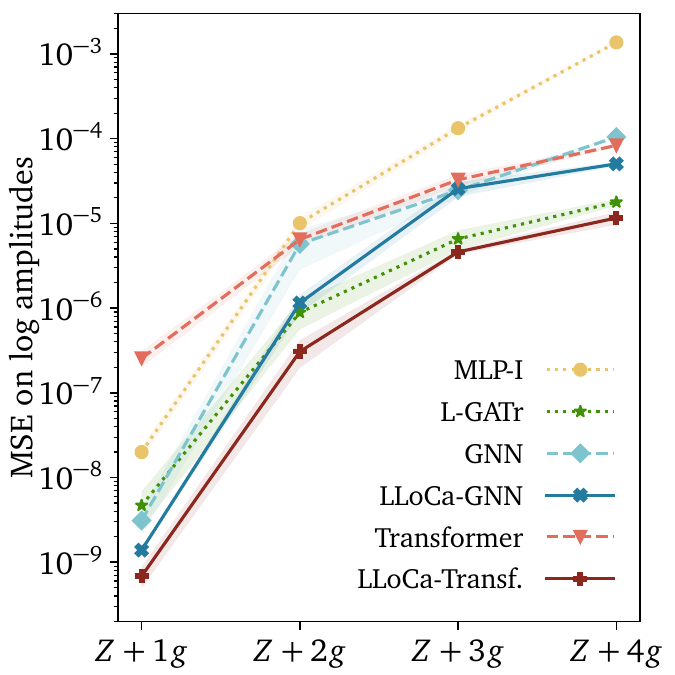}
    \end{minipage}
    \begin{minipage}{0.52\linewidth}
    \centering
    \begin{small}
    \begin{tabular}{l S[table-format=3.1] @{\ \error{$\pm$} \ } l @{\hskip 0.4cm} S[table-format=2.1] S[table-format=2.1]}
    \toprule
     Network  & \multicolumn{2}{c}{MSE$\times 10^{-5}$} & FLOPs & \text{Time} \\ \midrule
     MLP-I~\cite{Spinner:2024hjm} & 137.0 & \error{2} & 0.1M & 0.4h \\
     L-GATr~\cite{Spinner:2024hjm} & 1.8 & \error{0.2} & 528.0M & 8.3h \\
     \addlinespace[0.5ex]
     GNN & 10.5 & \error{0.2} & 20.7M & 0.9h \\
     LLoCa-GNN & \phantom{00}5.0 & \error{0.2} & 22.3M & 1.5h \\
     \addlinespace[0.5ex]
     Transformer & 8.3 & \error{0.3} & 14.9M & 1.3h \\
     LLoCa-Transformer & \phantom{00} 1.2 & \error{0.2} & 16.3M & 2.3h \\
     \bottomrule
    \end{tabular}
    \end{small}
    \end{minipage}
    \caption{Left: scaling of the prediction error with the multiplicity for all amplitude surrogates. Right: training cost on $Z+4g$ amplitudes. Networks are trained on $10^7$ events for all multiplicities. Error bands are based on the mean and standard deviation of three different random seeds. These results are also included in Ref.~\cite{Spinner:2025prg}.}
    \label{fig:amp_multiplicity}
\end{figure}
To compare the LLoCa performance to our baselines in Fig.~\ref{fig:amp_multiplicity}, we train amplitude surrogates for different multiplicities each using $10^7$ training events. The transformers and graph networks scale well to high multiplicity due to their permutation equivariance, while the MLP-I is worse by one order of magnitude. The LLoCa networks consistently outperform their non-equivariant counterparts, with a small gain for the LLoCa-GNN because of the Lorentz-invariant edge features already included in the GNN. The LLoCa-Transformer even surpasses the performance of the more expensive L-GATr network. 

Regarding computational cost, the LLoCa variants require more training time, but they benefit from faster training of the backbone architecture.
Compared to an equivariant architecture with specialized layers, L-GATr, our networks are at least four times faster. The computational overhead of LLoCa comes from the frame-to-frame conversions in Sec.~\ref{sec:message_passing} and from the construction of local frames in Sec.~\ref{sec:construct_frames}, in equal proportions.

\begin{figure}[t!]
    \includegraphics[width=0.495\linewidth]{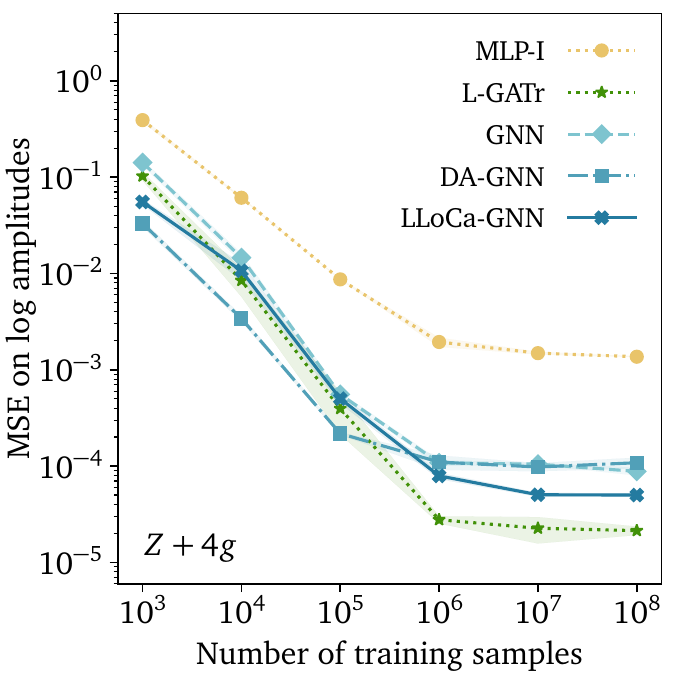}
    \includegraphics[width=0.495\linewidth]{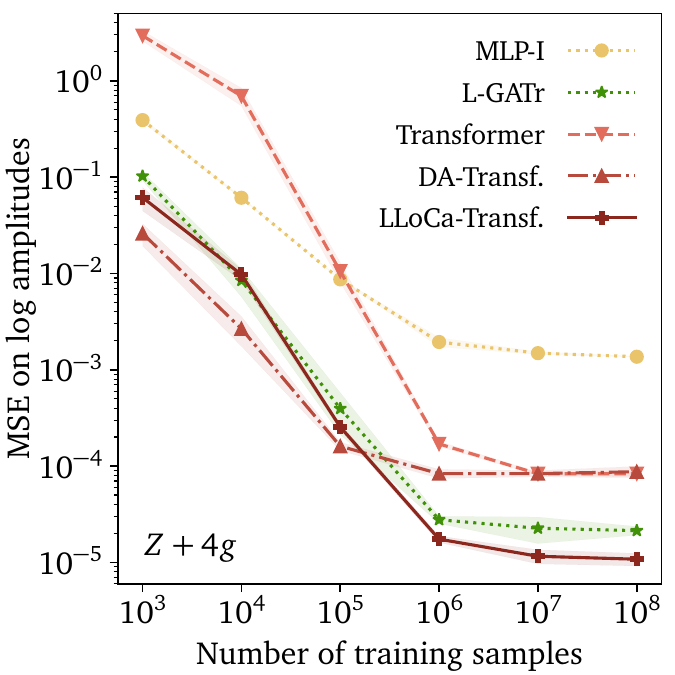}
    \caption{Exact Lorentz-equivariance vs.~data augmentation for GNNs (left) and transformers (right). Error bands are based on the mean and standard deviation of three different random seeds. These figures are also included in Ref.~\cite{Spinner:2025prg}.}
    \label{fig:ampxl_data}
\end{figure}

To understand the scaling of our amplitude surrogates with the training dataset size,  in Fig.~\ref{fig:ampxl_data} we focus on the process $Z+4g$ with the full $10^8$ training events. We compare the scaling of our LLoCa-GNN and LLoCa-Transformer with their non-equivariant versions as well as the MLP-I and L-GATr baselines. We also include networks trained with data augmentation, indicated by DA-GNN and DA-Transformer. More details on data augmentation for the Lorentz group are discussed in App.~\ref{app:training}.

Data augmentation emerges from LLoCa as a special case of random global frames, in contrast to the learned local frames of Sec.~\ref{sec:construct_frames}. Data augmentation, or learned equivariance, is a simple and cheap alternative to the exact Lorentz-equivariance encoded in our LLoCa networks. Networks trained with data augmentation match the corresponding baseline trainings for large data, where data augmentation eventually becomes redundant, and network capacity limits performance. The exactly equivariant LLoCa networks perform significantly better in this limit because they do not have to attribute parameters to learn the Lorentz-invariance of the amplitude. 
For less data, exact Lorentz-equivariance and data augmentation both show significantly better performance than non-equivariant networks, indicating that the symmetry information present in the architecture may prevent the network from overfitting. 

\begin{figure}[b!]
    \includegraphics[width=0.495\linewidth]{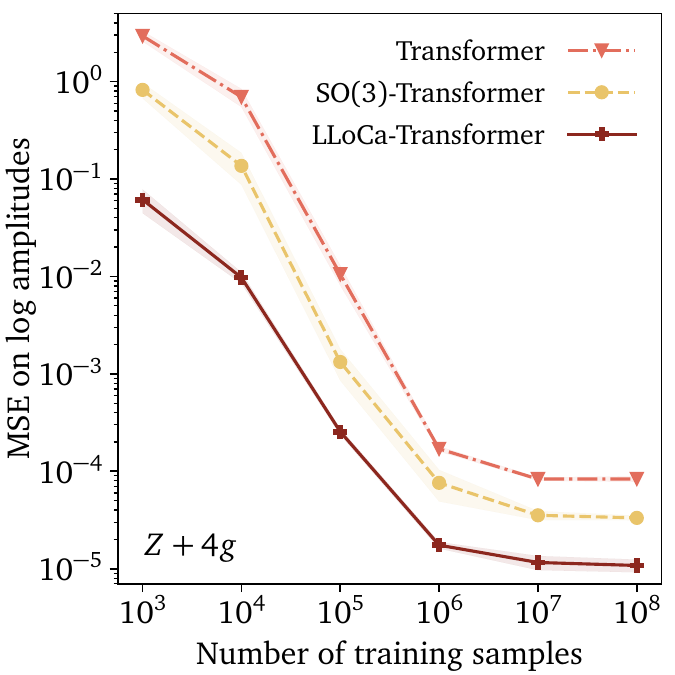}
    \includegraphics[width=0.495\linewidth]{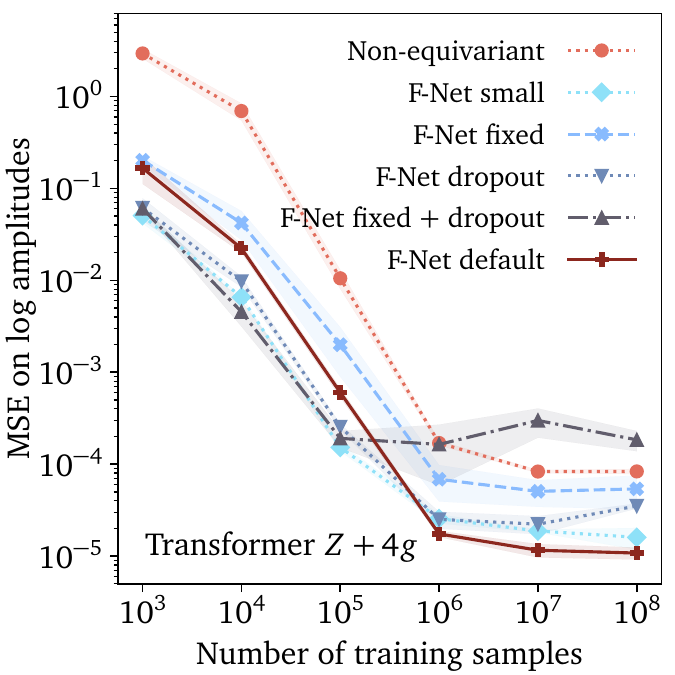}
    \caption{Effect of the choice of symmetry group (left) and options for the Frames-Net (right) on the LLoCa-Transformer. Error bands are based on the mean and standard deviation of three different random seeds. These figures are also included in Ref.~\cite{Spinner:2025prg}.}
    \label{fig:ampxl_abl_method}
\end{figure}

Focusing on the transformer, in Fig.~\ref{fig:ampxl_abl_method} we study the design choices of LLoCa more carefully. First, we test the SO(3)-Transformer, introduced in Sec.~\ref{sec:construct_frames} as a constrained version of the LLoCa-Transformer. For Lorentz-invariant regression targets, we indeed find that the fully Lorentz-equivariant transformer achieves the highest performance, whereas the SO(3)-Transformer is only slightly better than the non-equivariant baseline. In the right panel of Fig.~\ref{fig:ampxl_abl_method}, we study the effect of the Frames-Net that is used to construct the local frames. We find that decreasing the network size to only 1k learnable parameters slightly degrades performance in the large-data regime but improves performance in the small-data regime. This indicates that the Frames-Net overfits sooner than the backbone. To investigate this overfitting further, we included tests with fixed Frames-Net parameters and with a dropout probability of 0.2 in the Frames-Net. We find that even a fixed Frames-Net consistently improves performance over the non-equivariant baseline and that using dropout only in the Frames-Net helps to avoid overfitting.

We now investigate the effect of the hidden message representation $\rho_m$ in Tab.~\ref{tab:ampxl_abl_messages}. We find that LLoCa improves the performance compared to the non-equivariant baseline as long as any higher-order representation is included. LLoCa can even operate on complete rank-2 tensor representations, which is not possible in early Lorentz-equivariant architectures. We achieve the strongest performance by equally splitting hidden features in scalar and vector representations. Limiting LLoCa networks to scalar features only degrades the performance below the non-equivariant baseline because the network is forced to encode tensorial information but cannot meaningfully communicate it across particles. Finally, we implement global canonicalization by averaging the $v_k^\mu$ entering Eq.~\eqref{eq:gram-schmidt} over all particles, resulting in a learned global frame for the entire event. This global canonicalization achieves a performance halfway between local canonicalization and the non-equivariant baseline.

\begin{table}
    \centering
    \begin{small}
    \begin{tabular}{l S[table-format=2.1] @{\ \error{$\pm$} \ }  l }
    \toprule
     Method & \multicolumn{2}{c}{MSE ($\times 10^{-5}$)}  \\ 
     \midrule
     Non-equivariant & 8.3 & \error{0.5} \\
     Global canonical. & 4.4 & \error{1}  \\
     LLoCa ($16$ scalars) & 40 & \error{4}  \\
     LLoCa (single rank 2 tensor) & 2.0 & \error{0.4} \\ 
     LLoCa ($4$ vectors) & 2.0 & \error{0.2} \\
     LLoCa ($12$ scalars, $1$ vector) & 1.2 & \error{0.1} \\
     LLoCa ($8$ scalars, $2$ vectors) & 1.0 & \error{0.1} \\\bottomrule
    \end{tabular}
    \end{small}
    \caption{Effect of the hidden message representation $\rho_m$ in each of the 16-dimensional attention heads on the amplitude surrogate quality for LLoCa-Transformers trained on the full $Z+4g$ dataset. The uncertainties are estimated from three different random seeds. This table is also included in Ref.~\cite{Spinner:2025prg}.}
    \label{tab:ampxl_abl_messages}
\end{table}

Finally, we compare the performance and computational cost of the PELICAN and L-GATr Frames-Net architectures with those of the MLP Frames-Net and the non-equivariant baseline, as shown in Tab.~\ref{tab:alternative_framesnet_amp}. We find that all three architectures achieve comparable performance, with the PELICAN variant performing slightly better than the MLP and the L-GATr performing slightly worse. The computational costs are also similar, since for events with only seven particles the memory advantage of transformer-based architectures is not yet significant.

\begin{table}
\centering
        \begin{tabular}{l S[table-format=2.1] @{\ \error{$\pm$} \ } l S[table-format=1.1] S[table-format=2.1] S[table-format=1.2] S[table-format=1.2]}
        \toprule
         Frames-Net & \multicolumn{2}{c}{MSE ($\times 10^{-5}$)} & \text{Time} & \text{FLOPs} & \text{Memory} & \text{Parameters} \\
         \midrule
         Non-equivariant & 8.3 & \error{0.5} & 1.3h & 14.9M & 0.44G & 1.06M \\
         L-GATr & 1.3 & \error{0.1} & 3.2h & 16.9M & 0.83G & 1.07M\\
         PELICAN & 0.9 & \error{0.1} & 3.2h & 16.1M & 0.88G & 1.07M\\
         MLP (default) & 1.0 & \error{0.1} & 2.3h & 16.3M & 0.80G& 1.07M\\
         \bottomrule
        \end{tabular}
\caption{Performance and computational cost of alternative Frames-Net architectures for amplitude regression using the full $Z+4g$ dataset. Memory and time are inferred from full trainings on a H100 GPU, whereas the FLOPs are extracted for a forward pass on a single event. The uncertainties are estimated from three different random seeds.}
\label{tab:alternative_framesnet_amp}
\end{table}

\section{Event generation with LLoCa}
\label{sec:generation}

After demonstrating that LLoCa networks can learn to regress QFT amplitudes, we now use LLoCa generators to learn the normalized phase space density directly.\footnote{All results in this section are new compared to Ref.~\cite{Spinner:2025prg}.} This enables fast sampling from the phase space density without the usual rejection step. Such event-level generative networks are required for neural importance sampling~\cite{Gao:2020vdv,Bothmann:2020ywa,Heimel:2022wyj,Heimel:2023ngj}, end-to-end generation~\cite{Butter:2019cae,Butter:2021csz,Butter:2023fov}, generative unfolding~\cite{Bellagente:2019uyp,Bellagente:2020piv,Backes:2022sph,Shmakov:2023kjj,Diefenbacher:2023wec,Huetsch:2024quz} and optimal inference~\cite{Butter:2022vkj,Heimel:2023mvw, Bhattacharya:2025ave}. Again, learning the full phase space density to a precision that renders the network uncertainty negligible is the key challenge that calls for physics-inspired network designs. 

We focus on the production of two hadronically decaying top quarks in association with up to four extra jets~\cite{Brehmer:2024yqw}
\begin{align}
    pp\to t_h \bar t_h + nj,\qquad n=0...4\;,
\end{align}
We closely follow the setup used in Ref.~\cite{Brehmer:2024yqw} to study the L-GATr generator. 
The samples are generated with MadGraph~3.5.1~\cite{madgraph}. MadEvent~\cite{Alwall:2011uj} supplies the hard‑scattering matrix elements, Pythia~8~\cite{Sjostrand:2014zea} performs the parton showering, and Delphes~3~\cite{deFavereau:2013fsa} emulates the detector response. Jets are reconstructed using the anti‑$k_T$ algorithm with $R=0.4$ as implemented in FastJet~\cite{Cacciari:2011ma}. Pythia is run without multi‑parton interactions and with the default ATLAS detector card. We apply the phase space cuts
\begin{align}
 p_{T,j} > 22~\gev 
 \qquad 
 \Delta R_{jj} > 0.5 
 \qquad |\eta_j| < 5 \; ,
\end{align} 
and demand exactly two $b$‑tagged jets. Events are reconstructed with a $\chi^{2}$ algorithm~\cite{ATLAS:2020ccu}, and identical particles are ranked by their transverse momentum $p_T$. The $t\bar t + nj$ sample sizes follow the relative production rates, yielding 9.8M, 7.2M, 3.7M, 1.5M, and 480k events for $n = 0,\dots,4$.

\subsection*{Conditional Flow Matching}
We construct generative networks using conditional flow matching. Conditional flow matching generators are continuous normalizing flows~\cite{chen2018neural}, a class of generative networks that parametrize the transition $x(t)$ between a simple latent distribution $x_1\sim \pl (x_1)$ and a phase space distribution $x_0\sim \pd (x_0)$ as a differential equation
\begin{align}
    \frac{dx(t)}{dt} = v(x(t), t)\;.
\end{align}
The velocity field $v(x(t),t)$ is a neural network, trained to satisfy the boundary conditions $x(0)\sim \pd (x)$, $x(1)\sim \pl (x)$. Conditional flow matching~\cite{lipman2023flowmatchinggenerativemodeling, albergo2023stochastic} is a simple and efficient way to train continuous normalizing flows based on conditional target trajectories
\begin{align}
    x(t | x_0, x_1) = (1-t)x_0 + t x_1 \to \begin{cases}x_0 & t\to 0\\ x_1& t\to 1.\end{cases}
\end{align}
The unconditional velocity field $v(x(t), t)$ is then optimized to match the conditional target trajectories,
\begin{align}
    v((1-t)x_0 + tx_1, t) \approx x_1 - x_0.
\end{align}
With a optimized velocity field $v(x(t), t)$, we can sample from the phase space distribution implicitly defined above using a fast ODE solver
\begin{align}
    x_0 = x_1 - \int_0^1 dt\; v(x(t),t)\;.
\end{align}

For our event generation task, the velocity field $v(x(t), t)$ is a function of several particle four-momenta. This function is equivariant under global $\mathrm{SO}(2)$ rotations around the beam axis as a result of the detector geometry. We emphasize that the velocity field in conditional flow matching requires a $\mathrm{SO}(2)$-equivariant neural network, whereas amplitude regression only required invariant networks. Further, previous studies~\cite{Heimel:2023mvw, Huetsch:2024quz, Brehmer:2024yqw} found that permutation-equivariant architectures like transformers which include particle type information to break permutation equivariance significantly outperform simple MLPs. For this reason, we focus on two transformer architectures for the velocity field $v(x(t), t)$~\cite{Brehmer:2024yqw}
\begin{itemize}
    \item a standard transformer (and its LLoCa version);
    \item L-GATr, a Lorentz-equivariant transformer.
\end{itemize}
Both the L-GATr and LLoCa-Transformer velocity networks are fully Lorentz-equivariant, but the data only has a partial $\mathrm{SO}(2)$ symmetry. Following the description in Sec.~\ref{sec:symmetry_breaking}, we include reference vectors for the time and beam directions to break the equivariance group down to a residual $\mathrm{SO}(2)$. For LLoCa, we can alternatively break the Lorentz-equivariance at the architecture level. Finally, we note that some kind of Lorentz symmetry breaking is necessary for all Lorentz-equivariant generative networks, as it is not possible to construct a normalized density that is invariant under a non-compact group like the Lorentz group.

We follow Ref.~\cite{Brehmer:2024yqw} in our choice of base distribution $\pl (x_1)$, using gaussian distributions in the coordinates $(p_x,p_y,p_z,\log m^2)$ with mean and standard deviation fitted to the phase space distribution $\pd (x_0)$. This choice for $\pl (x_1)$ is invariant under the residual $\mathrm{SO}(2)$ group, which is necessary to construct a $\mathrm{SO}(2)$-equivariant generative network. Furthermore, we use rejection sampling to enforce the constraints $p_T>22\,\gev$, $\Delta R>0.5$ already at the level of the base distribution.

Ref.~\cite{Brehmer:2024yqw} found that the choice of target trajectory is crucial for the performance of CFM phase space generators. We follow their choice of target trajectory as straight lines in the parametrization $x$, defined as
\begin{align}
p =
\begin{pmatrix} 
 E\\p_x\\p_y\\p_z 
\end{pmatrix} 
\quad \to  \quad
f^{-1}(p) = x =  
\begin{pmatrix} 
 x_p  \\ x_m \\ x_\eta \\ x_\phi
\end{pmatrix} 
\equiv
\begin{pmatrix} 
 \log( p_T-p_T^\text{min}) \\ \log m^2 \\ \eta \\ \phi
 \end{pmatrix} \; .
\label{eq:momentum_rep}
\end{align}

For the Lorentz-equivariant L-GATr and LLoCa networks, we follow Ref.~\cite{Brehmer:2024yqw} to first map $x$ to $f(x)=p$, evaluate the networks with $p$ as input, extract the four-vector $v_p(p(t),t)$, and then extract $v_x (x(t),t)$ as
\begin{align}
    v_x(x(t),t) = \frac{\partial f^{-1}(p)}{\partial p} v_p(p(t),t)\;.
\end{align}
We also follow their setup for stabilizing the components $v_{x_p}$ and $v_{x_m}$ of the extracted velocity field $v_x$ by overwriting them directly with scalar outputs from the Lorentz-equivariant network. This step introduces an additional layer of symmetry-breaking from the Lorentz group to the residual $\mathrm{SO}(2)$.

For the LLoCa-Transformer, we find that large boosts, i.e.~a large boost factor $\gamma$ in the construction of the boost $B$ in Eq.~\eqref{eq:boost_parametrization}, affect the numerical stability of the network training. Specifically, large values of $\gamma$ cause large entries in the boost $B$, which are multiplied in each attention layer during the network forward pass, potentially leading to exploding gradients. The boost factors $\gamma$ average around $\gamma\approx 1.4$, but spikes up to $\gamma \approx 10$ can cause instabilities. We find that enforcing an upper limit $\gamma \le 3$ tames the instabilities and allows a stable training. We include a detailed discussion of this regularization procedure in App.~\ref{app:tests}.

\subsection*{Results}

\begin{figure}[t!]
    \includegraphics[width=0.33\textwidth,page=31]{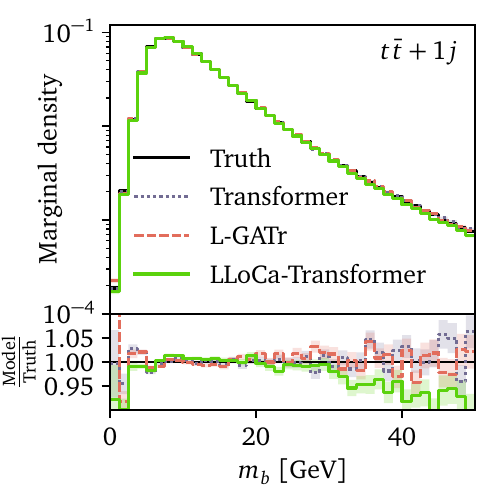}
    \includegraphics[width=0.33\textwidth,page=12]{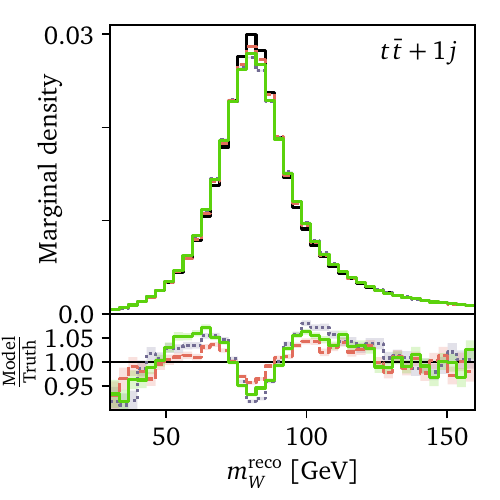}
    \includegraphics[width=0.33\textwidth,page=7]{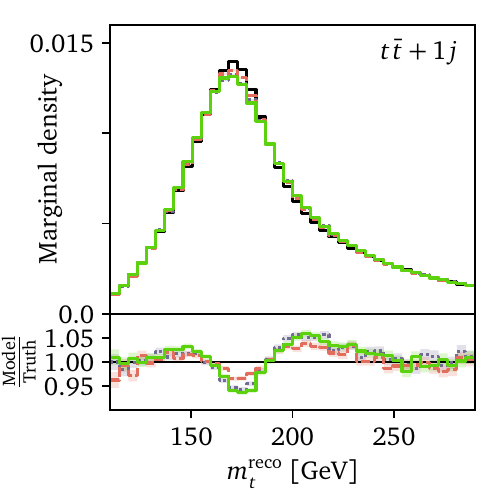} \\
    \includegraphics[width=0.33\textwidth,page=37]{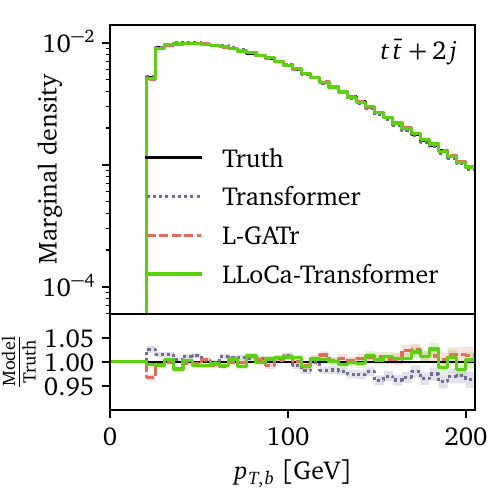}
    \includegraphics[width=0.33\textwidth,page=21]{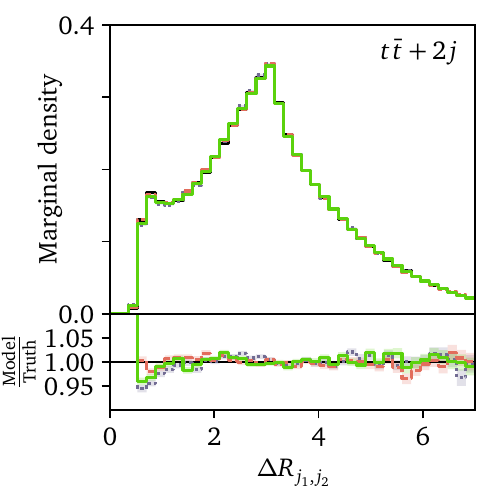}
    \includegraphics[width=0.33\textwidth,page=8]{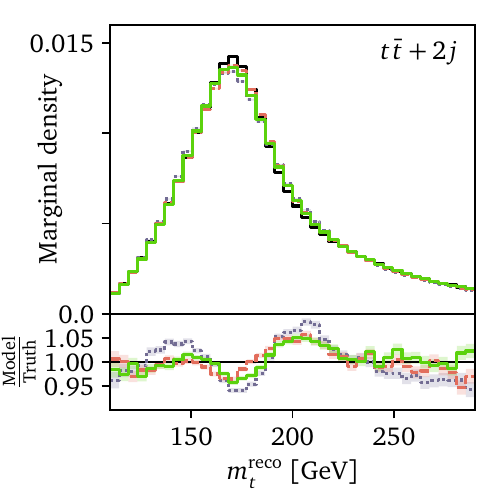} \\
    \includegraphics[width=0.33\textwidth,page=43]{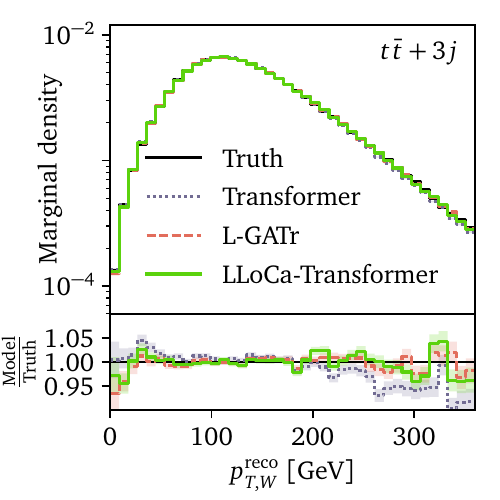}
    \includegraphics[width=0.33\textwidth,page=19]{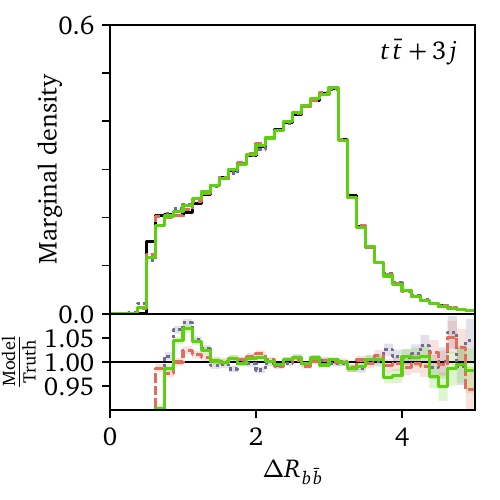}
    \includegraphics[width=0.33\textwidth,page=9]{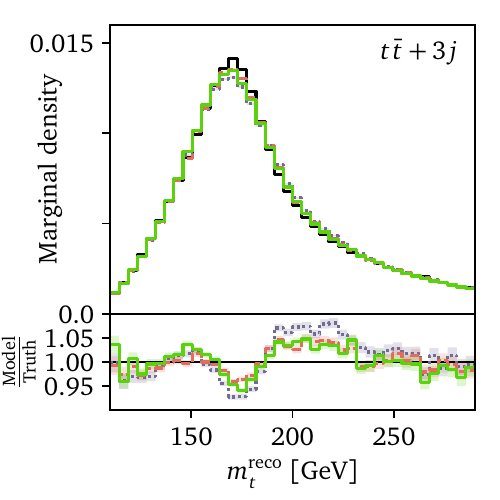} \\
    \includegraphics[width=0.33\textwidth,page=29]{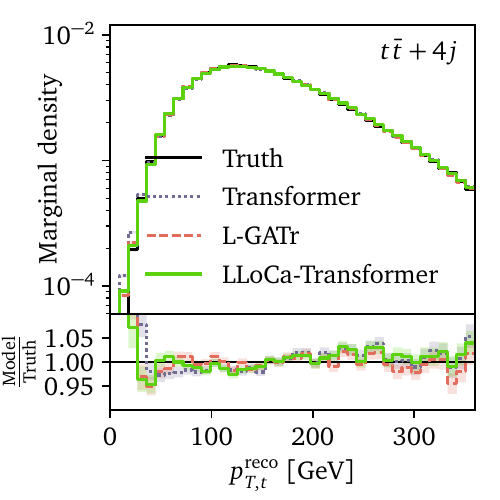}
    \includegraphics[width=0.33\textwidth,page=24]{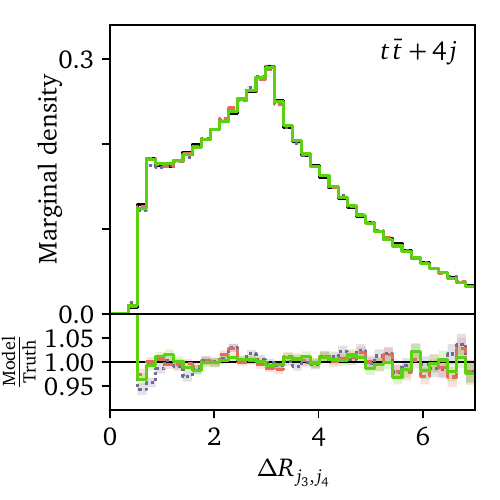}
    \includegraphics[width=0.33\textwidth,page=10]{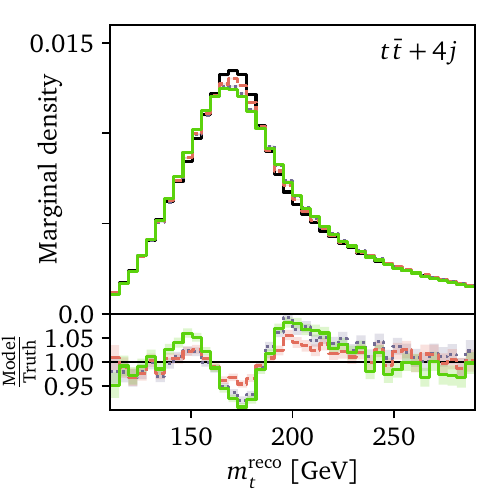}
    \caption{Overview of marginal distributions for $t\bar t+1,2,3,4$ jets (top to bottom).}
    \label{fig:eventgen-distributions}
\end{figure}

\begin{figure}[t!]
    \includegraphics[width=0.495\textwidth]{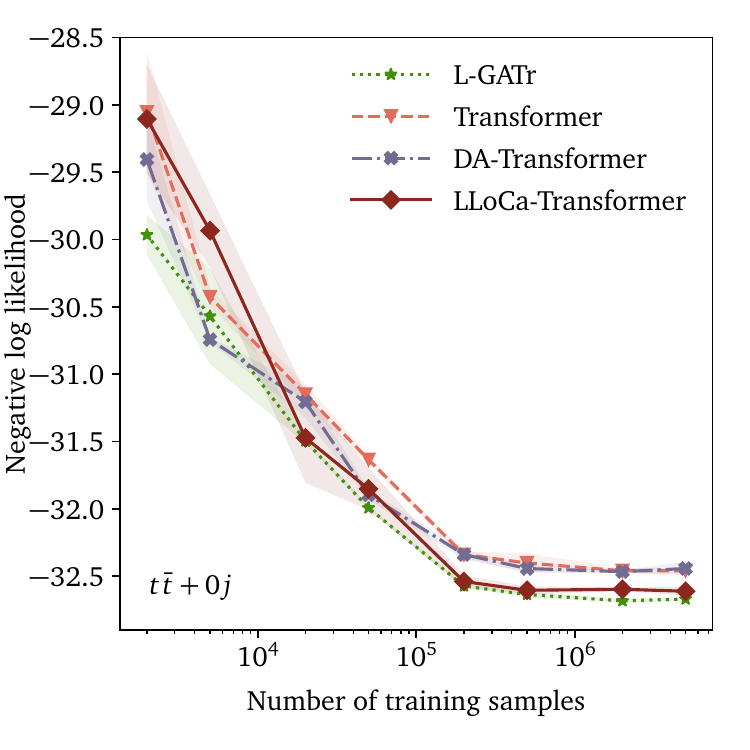}
    \includegraphics[width=0.495\textwidth]{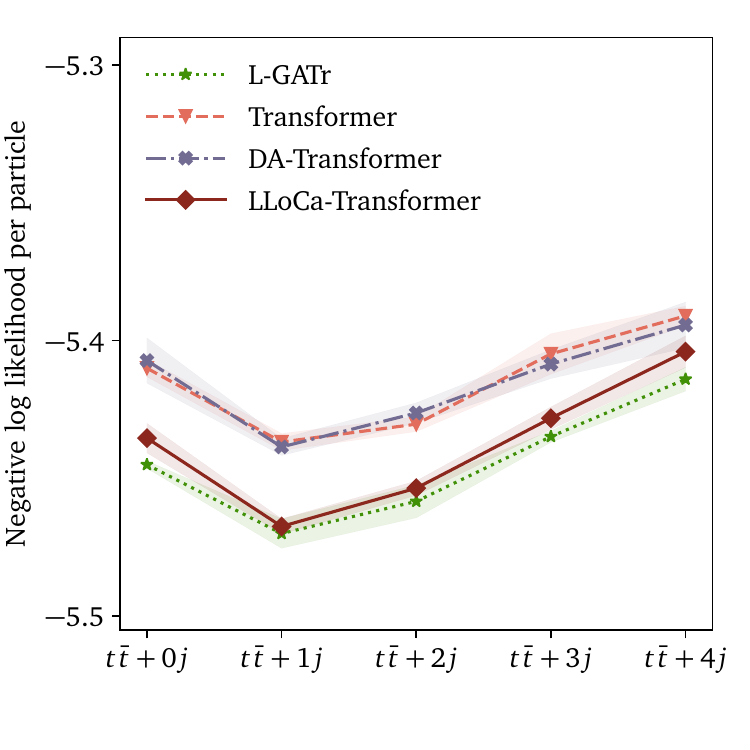}
    \includegraphics[width=0.495\textwidth]{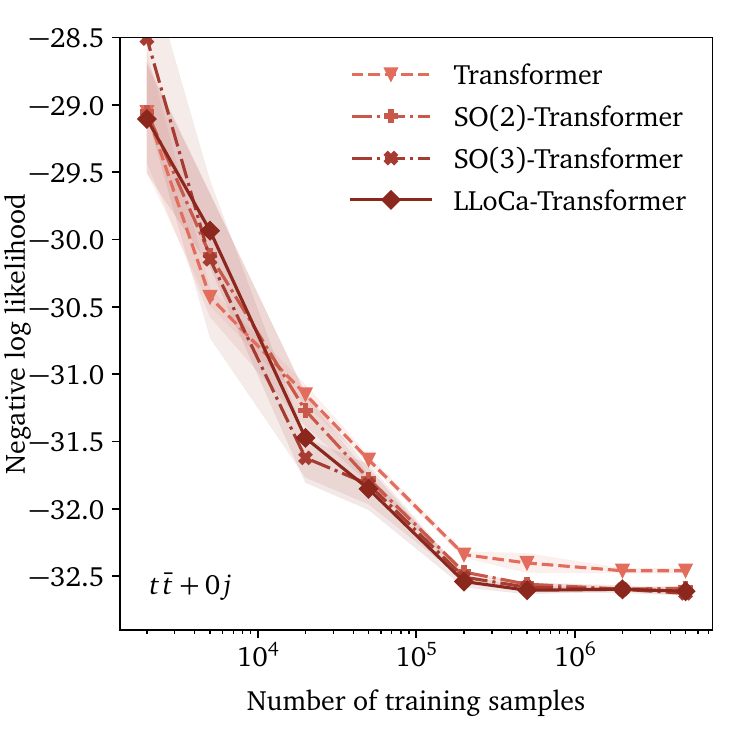}
    \includegraphics[width=0.495\textwidth]{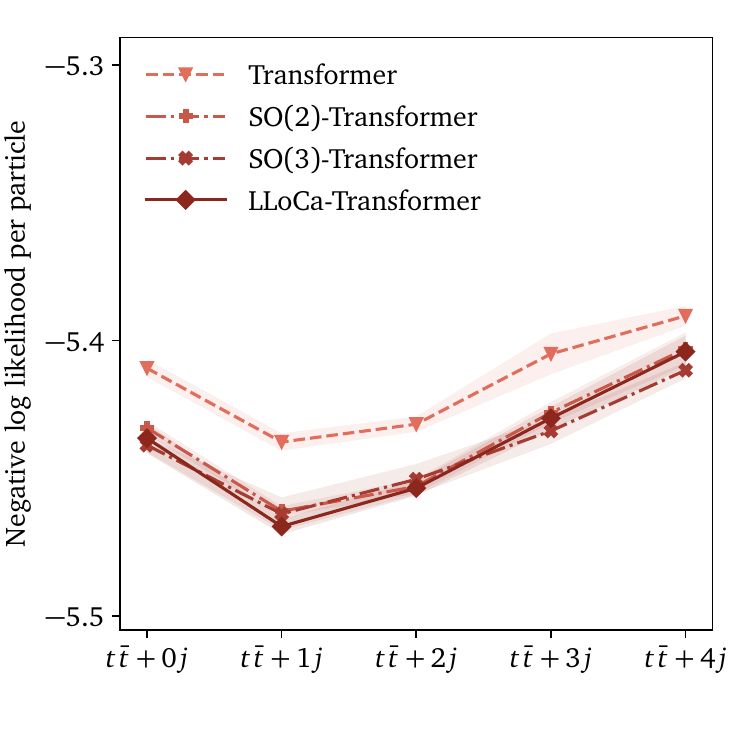}
    \caption{Negative log-likelihood of different generative networks across training dataset sizes (upper left) and event multiplicities (upper right). The lower panels show the negative log-likelihood of LLoCa-Transformers with varying degrees of symmetry breaking, again for different dataset sizes (lower left) and event multiplicities (lower right). Error bands represent the mean and standard deviation over three random seeds.}
    \label{fig:eventgen-scaling}
\end{figure}

To compare the performance of the baseline transformer with the LLoCa-Transformer and L-GATr, we study the selected kinematic distributions shown in Figure~\ref{fig:eventgen-distributions}. While all three architectures learn the phase space density to a comparable level and still struggle with a set of features, the Lorentz-equivariant LLoCa-Transformer and L-GATr networks show slightly better distributions than the standard transformer. All networks struggle to correctly learn the mass distributions of virtual particles, a scalar feature which remains challenging to learn even with Lorentz-equivariant networks.

In the upper panels of Figure~\ref{fig:eventgen-scaling}, we evaluate the negative log-likelihood on the test dataset, the standard scalar metric for generative networks.  Corresponding results using the neural classifier metric can be found in Appendix~\ref{app:generation-extra}. We show the scaling of the negative log-likelihood with the training dataset size and with the jet multiplicity. In addition to the transformer, LLoCa-Transformer and L-GATr networks, we include a generative network trained with data augmentation as DA-Transformer similar to how we did it for amplitude regression in Figure~\ref{fig:ampxl_data}. With this global metric at hand, we see that the Lorentz-equivariant LLoCa-Transformer and L-GATr networks consistently outperform the non-equivariant transformer and DA-Transformer when sufficient training data is available. This is in agreement with the findings of Section~\ref{sec:amplitudes}, where we concluded that Lorentz-equivariant networks win at scale due to the more efficient allocation of network parameters. In the small-data regime, the uncertainty from the training process estimated from repeated trainings is of similar magnitude as the differences between networks, making the non-equivariant transformer as the cheapest option a viable choice.

\begin{table}[t!]
\centering
\begin{small}
\begin{tabular}{lll}
\toprule
Network & NLL & AUC \\
\midrule
Transformer & \result{-32.46}{\dz{}0.02} & \result{0.567}{\dz{}0.002} \\
LLoCa-Transformer (13 scalars) & \result{-29.50}{\dz{}0.18} & \result{0.892}{\dz{}0.003} \\
LLoCa-Transformer (9 scalars, 1 vector) & \result{-32.61}{\dz{}0.02} & \result{0.509}{\dz{}0.001} \\
LLoCa-Transformer (5 scalars, 2 vectors) & \result{-32.61}{\dz{}0.03} & \result{0.508}{\dz{}0.001} \\
\bottomrule
\end{tabular}
\end{small}
\caption{Effect of the latent feature representation for the LLoCa-Transformer generator with a 13-dimensional attention head trained on the $t\bar t+0j$ dataset. We show the negative log-likelihood (NLL) metric as well as the AUC of a neural classifier, see Appendix~\ref{app:generation-extra} for more results on the neural classifier metric. Uncertainties are estimated from three random seeds.}
\label{tab:eventgen_representations}
\end{table}

In the lower panels of Figure~\ref{fig:eventgen-scaling}, we study the effect of symmetry breaking at the architecture level as introduced in Section~\ref{sec:symmetry_breaking} using the negative log-likelihood metric. The SO(2)-Transformer, SO(3)-Transformer and LLoCa-Transformer are all effectively only SO(2)-invariant, but for SO(3)-Transformer and LLoCa-Transformer reference vectors are used to break Lorentz-equivariance at the network input level. We find that the performance of these three designs agrees within uncertainties, and all are significantly better than the standard transformer. We conclude that for event generation the LLoCa networks do not make use of the option to restore full Lorentz-equivariance in some phase space regions, even though the reference vectors formally allow it. An SO(2)-equivariant architecture is therefore sufficient for this task. Being able to draw these conclusions is a benefit of the LLoCa framework which allows the user to easily modify the equivariance group. Such comparisons are more difficult for Lorentz-equivariant architectures relying on specialized layers, because the whole architecture has to be rewritten to modify the equivariance group. A previous study comparing the Lorentz-equivariant L-GATr and the E(3)-equivariant E(3)-GATr~\cite{Brehmer:2024yqw} found that E(3)-equivariance is on par with non-equivariance, and concluded that SO(3)-equivariance has no benefits for event generation. We suspect that the performance drop from L-GATr to E(3)-GATr was instead caused by the unused translation representations in E(3)-GATr, which do not allow for a fair comparison of L-GATr and E(3)-GATr at equal parameter count.

Finally, we study the impact of latent feature representations on the LLoCa-Transformer performance in Table~\ref{tab:eventgen_representations}. We find the same hierarchy as for amplitude regression, with latent feature representations including at least one vector being significantly better than scalar-only latent feature representations.

\section{Jet tagging with LLoCa}
\label{sec:tagging}

Our last application of the LLoCa networks is jet tagging, the prime testing ground where novel architectures have to prove their performance.\footnote{This section extends the results from Ref.~\cite{Spinner:2025prg} by the top tagging and TopTagXL datasets, as well as the PELICAN-lite network and studies on the impact of symmetry breaking in the LLoCa framework.} A series of Lorentz-equivariant networks~\cite{Gong:2022lye,Bogatskiy:2023nnw,ruhe2023clifford,Brehmer:2024yqw} shows the strongest performance on small-scale datasets, whereas transformers~\cite{Qu:2022mxj,Wu:2024thh,Brehmer:2024yqw,Krause:2025qnl,Rai:2025cog} are necessary to scale to large-scale datasets. Transformers with a focus on learnable edge features can match the performance of Lorentz-equivariant networks on small datasets~\cite{Wu:2024thh,Esmail:2025kii}, but currently fail to carry these gains to larger datasets. Only fully Lorentz-equivariant transformers significantly improve performance on large-scale datasets~\cite{Brehmer:2024yqw}. Finally, pretrained transformers with and without Lorentz-equivariance can outperform Lorentz-equivariant baselines on small datasets~\cite{Qu:2022mxj,Brehmer:2024yqw}. 

In this section, we create LLoCa versions of established non-equivariant jet taggers and compare them to the simple LLoCa-Transformer discussed in the previous sections. We compare our LLoCa architectures with literature baselines on the established top tagging and JetClass datasets and establish a larger benchmark dataset for top tagging, the TopTagXL dataset. 

We compare LLoCa to a set of cutting-edge jet taggers, to allow a direct comparison on the JetClass and TopTagXL datasets:
\begin{itemize}
    \item ParticleNet~\cite{Qu:2019gqs}, a convolutional graph neural network (and its LLoCa version);
    \item ParT~\cite{Qu:2022mxj}, a transformer with learnable attention bias based on edge features (and its LLoCa version);
    \item LorentzNet~\cite{Gong:2022lye}, a Lorentz-equivariant graph network;
    \item PELICAN-lite, an efficient reimplementation of the Lorentz-invariant graph network PELICAN~\cite{Bogatskiy:2023nnw} described in App.~\ref{app:pelican};
    \item L-GATr~\cite{Brehmer:2024yqw}, a Lorentz-equivariant transformer;
    \item MIParT~\cite{Wu:2024thh}, an extension of ParT with more operations on edge features;
    \item a standard transformer (and its LLoCa version).
\end{itemize}
We compare the computational cost in terms of timing, memory usage, FLOPs and parameter count for all networks in Table~\ref{tab:jetclass_tagging}.

\begin{table}[t!]
\centering
\begin{small}
\begin{tabular}{lll S[table-format=3.0] @{\hspace{0.4cm}} S[table-format=4.0] @{\hspace{0.4cm}} S[table-format=2.1] @{\hspace{0.4cm}} S[table-format=3.0]}
\toprule
Network & Accuracy & AUC & \text{Time} & \text{FLOPs} & \text{Memory} & \text{Parameters} \\
\midrule
MIParT-L \cite{Wu:2024thh} & 0.860\dz{} & 0.9877\dz{} & 43h & 225M & 53.6G & 2380k \\
LorentzNet* \cite{Gong:2022lye} & 0.849\dz{} & 0.9858\dz{} & 64h & 676M & 20.5G & 223k \\
PELICAN-lite* & 0.856\dz{} & 0.9869\dz{} & 97h & 1370M & 27.4G & 244k \\
L-GATr* \cite{Brehmer:2024yqw} & 0.865\dz{} & 0.9884\dz{} & 166h & 2060M & 19.0G & 1079k \\
\addlinespace[0.5ex]
ParticleNet \cite{Qu:2019gqs} & 0.845\dz{} & 0.9851\dz{} & 25h & 413M & 16.5G & 366k \\
LLoCa-ParticleNet*  & 0.849 & 0.9858 & 43h & 517M & 23.5G & 385k \\
\addlinespace[0.5ex]
ParT \cite{Qu:2022mxj} & 0.861\dz{} & 0.9878\dz{} & 33h & 211M & 13.3G & 2141k \\
LLoCa-ParT* & 0.864 & 0.9882 & 66h & 315M & 19.9G & 2160k \\
\addlinespace[0.5ex]
Transformer & 0.855 & 0.9867 & 15h & 210M & 2.3G & 1979k \\
LLoCa-Transformer* & 0.864 & 0.9882  & 31h & 301M & 6.9G & 1998k \\
\bottomrule
\end{tabular}
\end{small}
\caption{Performance and computational cost for taggers on the JetClass dataset. We show accuracy, AUC, time for a complete training on a H100 GPU, FLOPs per forward pass, maximum memory consumption during training, and number of learnable parameters. Note that the results for MIParT-L, L-GATr, ParticleNet and ParT differ slightly from the original publications because we re-trained the networks. Uncertainties in accuracy and AUC are estimated based on 5 independent seeds and at most $\pm 1$ in the last significant digit, covering the differences to the results reported in the original publications. Lorentz-equivariant networks are denoted with an asterisk. See Table~\ref{tab:jetclass_rejection} for background rejection rates at fixed signal efficiency. The LLoCa tagger results in this table were first presented in Ref.~\cite{Spinner:2025prg}.}
\label{tab:jetclass_tagging}
\end{table}

\begin{table}[t]
    \fontsize{9}{9}\selectfont
    \addtolength{\tabcolsep}{-0.1em}
    \centering
    \begin{tabular}{l ccccc}
        \toprule
        & $H \to b\bar{b}$ & $H \to c\bar{c}$ & $H \to gg$ & $H \to 4q$ & $H \to l\nu q\bar{q}'$ \\        
        & Rej$_{50\%}$ & Rej$_{50\%}$ & Rej$_{50\%}$ & Rej$_{50\%}$ & Rej$_{99\%}$ \\
        \midrule
        MIParT-L~\cite{Wu:2024thh} & \result{10532}{239}& \result{4304}{55}& \result{121}{1}& \result{1920}{33}& \result{5434}{216}\\
        LorentzNet*~\cite{Gong:2022lye} & \result{\dz{}8722}{160}& \result{2844}{24}& \result{113}{0}& \result{1143}{21}& \result{3564}{\dz{}85}\\
        PELICAN-lite* & \result{\dz{}8869}{301}& \result{3586}{71}& \result{117}{0}& \result{1494}{27}& \result{3987}{\dz{}30}\\
        L-GATr*~\cite{Brehmer:2024yqw} & \result{11759}{308}& \result{4741}{89}& \result{128}{0}& \result{2197}{54}& \result{5861}{120}\\
\addlinespace[0.5ex]
        ParticleNet~\cite{Qu:2019gqs} & \result{\dz{}8084}{396}& \result{2600}{62}& \result{106}{0}& \result{\dz{}957}{12}& \result{3160}{\dz{}74}\\
        LLoCa-ParticleNet* & \result{\dz{}7813}{230}& \result{3005}{49}& \result{107}{0}& \result{1121}{17}& \result{3267}{\dz{}80}\\
\addlinespace[0.5ex]
        ParT~\cite{Qu:2022mxj} & \result{11104}{446}& \result{4354}{70}& \result{123}{0}& \result{1891}{14}& \result{5465}{\dz{}66}\\
        LLoCa-ParT* & \result{11909}{441}& \result{4578}{82}& \result{125}{1}& \result{2045}{33}& \result{5635}{151}\\
\addlinespace[0.5ex]
        Transformer & \result{10432}{224}& \result{3456}{34}& \result{116}{0}& \result{1371}{10}& \result{4632}{115}\\
        LLoCa-Transformer* & \result{11704}{310}& \result{4626}{54}& \result{125}{1}& \result{2086}{20}& \result{5828}{\dz{}58}\\
        \bottomrule
    \end{tabular}
    \begin{tabular}{l cccc}
        \toprule
        & $t \to b q\bar{q}'$ & $t \to b l\nu$ & $W \to q\bar{q}'$ & $Z \to q\bar{q}$ \\
        & Rej$_{50\%}$ & Rej$_{99.5\%}$ & Rej$_{50\%}$ & Rej$_{50\%}$ \\
        \midrule
        MIParT-L~\cite{Wu:2024thh} & \result{31522}{1555}& \result{17026}{917}& \result{535}{5}& \result{397}{3}\\
        LorentzNet*~\cite{Gong:2022lye} & \result{14709}{\dz{}325}& \result{11379}{787}& \result{389}{1}& \result{310}{1}\\
        PELICAN-lite* & \result{19054}{\dz{}919}& \result{12199}{488}& \result{470}{2}& \result{349}{2}\\
        L-GATr*~\cite{Brehmer:2024yqw} & \result{39326}{2897}& \result{18610}{642}& \result{574}{4}& \result{427}{4}\\
\addlinespace[0.5ex]
        ParticleNet~\cite{Qu:2019gqs} & \result{10299}{\dz{}336}& \result{10424}{548}& \result{352}{2}& \result{286}{2}\\
        LLoCa-ParticleNet* & \result{11109}{\dz{}326}& \result{\dz{}9856}{364}& \result{420}{3}& \result{317}{1}\\
\addlinespace[0.5ex]
        ParT~\cite{Qu:2022mxj} & \result{30762}{\dz{}542}& \result{16497}{801}& \result{543}{4}& \result{399}{2}\\
        LLoCa-ParT* & \result{36456}{2901}& \result{18086}{786}& \result{555}{4}& \result{413}{5}\\
\addlinespace[0.5ex]
        Transformer & \result{25846}{1363}& \result{16562}{316}& \result{412}{3}& \result{334}{1}\\
        LLoCa-Transformer* & \result{37459}{2265}& \result{18530}{452}& \result{553}{5}& \result{414}{3}\\
        \bottomrule
    \end{tabular}
    \vspace{0.2cm}
    \caption{Background rejection rates $1/\epsilon_B$ for the JetClass dataset~\cite{Qu:2022mxj}. For each class, $\mathrm{Rej}_{N\%}$ represents the inverse signal fraction at fixed background rejection rate $N\%$. Lorentz-equivariant methods are marked with an asterisk*. Similar to Figure~\ref{tab:jetclass_tagging}, uncertainties are based on trainings with five independent seeds, and all networks are re-trained independently.}
    \label{tab:jetclass_rejection}
\end{table}

\begin{figure}[t!]
    \centering
    \includegraphics[width=0.495\linewidth,page=6]{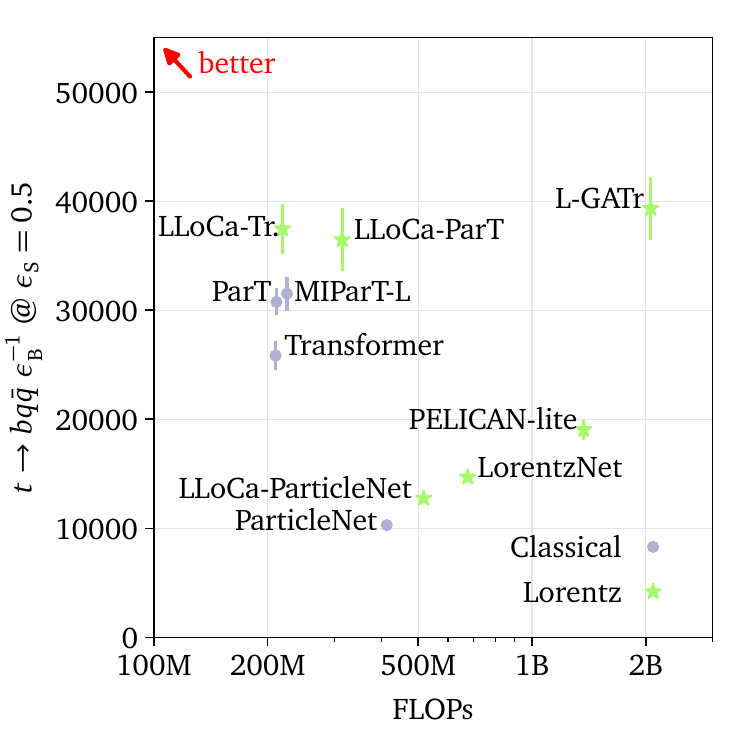}
    \includegraphics[width=0.495\linewidth,page=7]{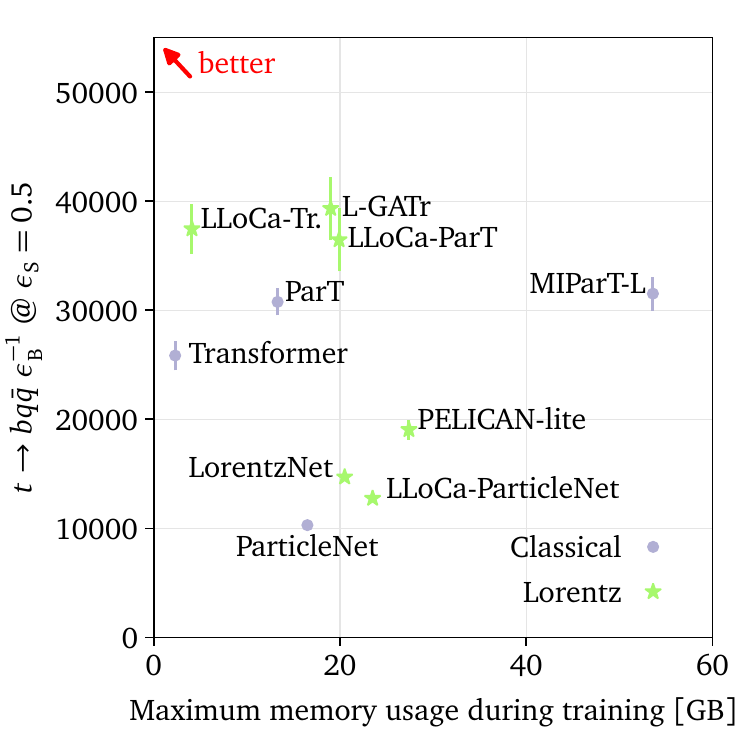}
    \caption{Performance vs FLOPs (left) and memory usage (right) for top-performing jet taggers on the JetClass dataset. These figures are a visual representation of the results presented in Tables~\ref{tab:jetclass_tagging},~\ref{tab:jetclass_rejection}.}
    \label{fig:jetclass_landscape}
\end{figure}

\subsection*{Multi-class tagging on JetClass}
The JetClass dataset~\cite{Qu:2022mxj} includes 10 classes with 10M events each, totaling 100M training jets.
The QCD class contains quark or gluons initiated jets, while the signals contain the production of $Z$, $W$, and $H$ bosons, and top-initiated jets.
The events are generated using Madgraph, followed by Pythia8 for showering and hadronization, and the detector interaction is simulated through Delphes using the CMS detector card.
The reconstructed jets satisfy
\begin{align}
p_{T,j} = 500,\ldots,1000\,\gev \qquad \text{and} \qquad |\eta_j| < 2.0
\end{align}
Jets are clustered using FastJet with the anti-$k_T$ algorithm with $R=0.8$.
For the signal jet classes, only fully reconstructed signatures are stored.
For each jet, the dataset provides the four-momenta of the constituents, kinematic jet features, particle identification numbers (PIDs), and trajectory displacement variables. 

We evaluate taggers in terms of accuracy, AUC score, training time on a single H100 GPU, number of floating point operations (FLOPs) for a single forward pass, maximum GPU memory used during training, and number of network parameters.
Table~\ref{tab:jetclass_tagging} and Table~\ref{tab:jetclass_rejection} contain the evaluation metrics for several equivariant and non-equivariant taggers, and Figure~\ref{fig:jetclass_landscape} provides a visual representation of the same information. Figure~\ref{fig:jetclass_landscape2} in Appendix~\ref{app:tagging-extra} contains a version of Figure~\ref{fig:jetclass_landscape} for the class-averaged AUC score.
We note that all the top-performing networks are Lorentz-equivariant. In particular, our LLoCa versions of ParticleNet and ParT improve over the respective baselines, demonstrating the advantage of including Lorentz-equivariance in jet tagging.
Moving from the simple transformer to the LLoCa-Transformer, we observe the largest gain in tagging performance. ParT, another transformer network, includes specific Lorentz-invariant features in the architecture. Both LLoCa-Transformer and LLoCa-ParT achieve an AUC of 0.9882 and accuracy of 0.864, demonstrating that the additional features in ParT are not needed if the entire network is Lorentz-equivariant.
In terms of computational cost, the LLoCa-Transformer achieves superior performance to ParT at slightly reduced training time. It also achieves performance comparable to L-GATr while being four times faster and showing a similar reduction in FLOPs, regardless of the additional number of parameters.

We also demonstrate that the local canonicalization and the tensorial latent representations are essential elements of the LLoCa architecture. 
In Tab.~\ref{tab:jetclass_tagging_representations}, we compare different choices for the internal representations of the network. As discussed in the previous sections, using only scalar representations fails to communicate tensorial information and is therefore not a viable option. The global canonicalization only matches the ParT performance, while the best performing option is a combination of both scalar and vectorial channels.

\begin{table}[t!]
\centering
\begin{small}
\begin{tabular}{lll}
\toprule
Network & Accuracy & AUC \\
\midrule
Transformer & 0.855 & 0.9867 \\
ParT & 0.861 & 0.9877 \\
LLoCa-Transformer (global can.) & 0.861 & 0.9878 \\
LLoCa-Transformer (16 scalars) & 0.851 & 0.9862 \\
LLoCa-Transformer (4 vectors) & 0.863 & 0.9880 \\
LLoCa-Transformer (8 scalars, 2 vectors) & 0.864 & 0.9882 \\
LLoCa-Transformer (12 scalars, 1 vector) & 0.864 & 0.9882 \\
\bottomrule
\end{tabular}
\end{small}
\caption{Effect of the choice of internal representation for a LLoCa-Transformer with 16-dimensional attention heads on the JetClass dataset, similar to Table~\ref{tab:ampxl_abl_messages} for amplitude regression.}
\label{tab:jetclass_tagging_representations}
\end{table}

\begin{table}
    \centering
    \begin{tabular}{l ll S[table-format=3.0] @{\hspace{0.4cm}} S[table-format=4.0] @{\hspace{0.4cm}} S[table-format=2.1] @{\hspace{0.4cm}} S[table-format=3.0]}
    \toprule
    Frames-Net & Accuracy & AUC & \text{Time} & \text{FLOPs} & \text{Memory} & \text{Parameters} \\
    \midrule
    Non-equivariant & 0.855\dz{} & 0.9867\dz{} & 15h & 210M & 2.3G & 1979k \\
    L-GATr & 0.862 & 0.9879 & 30h & 218M & 3.2G & 1986k \\
    PELICAN & 0.862 & 0.9879 & 37h & 288M & 7.2G & 1989k \\
    MLP (default) & 0.864\dz{} & 0.9882\dz{} & 31h & 301M & 6.9G & 1998k \\
    \bottomrule
    \end{tabular}
    \caption{Performance and computational cost of alternative Frames-Net architectures for jet tagging on the JetClass dataset. Memory and time are inferred from full trainings on a H100 GPU, whereas the FLOPs are extracted for a forward pass on a single event with 50 particles. All results are for the (LLoCa-)Transformer.}
    \label{tab:alternative_framesnet_tag}
\end{table}

Next, we compare the performance of different Frames-Net architectures introduced in Section~\ref{sec:framesnet-alternatives} in Tab.~\ref{tab:alternative_framesnet_tag}. Since this task involves events with up to 128 particles, the MLP Frames-Net incurs substantial memory overhead, increasing memory usage from 2.3GB to 6.2GB relative to the non-equivariant transformer baseline. By avoiding learnable edge operations, the L-GATr Frames-Net scales more favorably with a memory consumption of only 3.2GB and increased FLOPs by just 4\%. The PELICAN Frames-Net introduces even higher computational cost than the MLP variant. In terms of tagging performance, both the L-GATr and PELICAN Frames-Nets perform slightly worse than the MLP Frames-Net. We attribute this gap to architectural details in Eqs.~\eqref{eq:construct_v2} and~\eqref{eq:construct_v3} that slightly destabilize training, and expect that further design refinements should close it.

\subsection*{Top tagging}

\begin{table}[t!]
    \centering
    \footnotesize
    \begin{tabular}{l c llll}
        \toprule
        Network && Accuracy & AUC & $1/\epsilon_B$ ($\epsilon_S=0.5$) & $1/\epsilon_B$ ($\epsilon_S=0.3$) \\
        \midrule
        PFN \cite{Komiske:2018cqr} && 0.932\dz{} & 0.9819 & \result{247}{\dz{}3} & \dz{}\result{888}{\dz{}17} \\
        ParticleNet \cite{Qu:2019gqs} && 0.940\dz{} & 0.9858 & \result{397}{\dz{}7} & \result{1615}{\dz{}93} \\
        ParT \cite{Qu:2022mxj} && 0.940\dz{} & 0.9858 & \result{413}{\dz{}16} & \result{1602}{\dz{}81} \\
        MIParT \cite{Wu:2024thh} && 0.942\dz{} & 0.9868 & \result{505}{\dz{}8} & \result{2010}{\dz{}97} \\
        IAFormer \cite{Esmail:2025kii} && 0.942 & 0.987 & \result{510}{\dz{}6} & \result{2012}{\dz{}30} \\
        LorentzNet* \cite{Gong:2022lye} && 0.942\dz{} & 0.9868 & \result{498}{\dz{}18} & \result{2195}{\dz{}173} \\
        CGENN* \cite{ruhe2023clifford} &&  0.942\dz{} & 0.9869 & 500 & 2172 \\
        PELICAN* \cite{Bogatskiy:2023nnw} && \result{0.9426}{\dz{}0.0002} & \result{0.9870}{\dz{}0.0001} & -- & \result{2250}{\dz{}75} \\
        L-GATr* \cite{Spinner:2024hjm} && \result{0.9423}{\dz{}0.0002} & \result{0.9870}{\dz{}0.0001} & \result{540}{\dz{}20} & \result{2240}{\dz{}70} \\
        Transformer && \result{0.9393}{\dz{}0.0002} & \result{0.9855}{\dz{}0.0001} & \result{389}{\dz{}6} & \result{1613}{\dz{}118} \\
        LLoCa-Transformer* && \result{0.9416}{\dz{}0.0001} & \result{0.9866}{\dz{}0.0001} & \result{492}{\dz{}15} & \result{2150}{\dz{}130} \\
        \bottomrule
    \end{tabular}
    \caption{Top tagging accuracy, AUC and background rejection rates at fixed signal efficiency on the standard dataset. Lorentz-equivariant methods are indicated with an asterisk. Uncertainties are estimated using five random seeds.}
    \label{tab:top_tagging}
\end{table}

Next, we benchmark LLoCa on the established top tagging reference dataset~\cite{Kasieczka:2019dbj}. It includes jets from light quarks and hadronic tops divided in 1.2M events for training and two sets of 0.4M events for testing and validation. The simulation uses Pythia8~\cite{Sjostrand:2014zea} for the generation, showering, and hadronization, while Delphes~\cite{deFavereau:2013fsa} with the ATLAS default detector card performs the detector simulation.
Reconstructed jets have the kinematic constraints
\begin{align}
p_{T,j} = 550,\ldots,650\,\gev \qquad \text{and} \qquad |\eta_j| < 2.0 \;,
\end{align}
We display the performance of our transformer and LLoCa-Transformer compared to other literature results in Table~\ref{tab:top_tagging}. Our standard transformer almost matches the performance of ParticleNet and ParT. The LLoCa-Transformer clearly outperforms the non-equivariant ParticleNet and ParT, and almost matches the performance other Lorentz-equivariant networks such as LorentzNet, PELICAN and L-GATr. We found that the network size and regularization parameters are essential for good performance on this dataset, and expect that more tuning in the future allows LLoCa-Transformer to fully match the performance of other Lorentz-equivariant top taggers. Still, LLoCa-Transformer is faster to evaluate and less memory-intensive than all other Lorentz-equivariant taggers at fixed batch size, see Table~\ref{tab:jetclass_tagging}.

\begin{figure}[t!]
    \centering
    \includegraphics[width=0.495\linewidth]{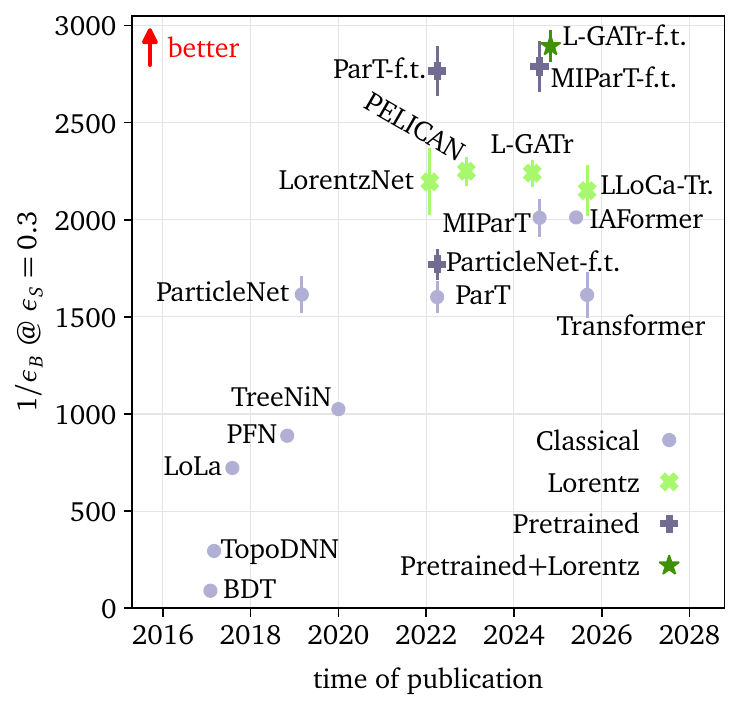}
    \includegraphics[width=0.495\linewidth,page=5]{figs/jetclass_landscape.pdf}
    \caption{Left: Progress in top tagging through advanced network architectures over time. Right: Efficiency of tagging architectures on the JetClass dataset. All networks are trained using the same batch size and number of iterations, leading to a roughly linear relationship between training and inference time.}
    \label{fig:money}
\end{figure}

The progress in top tagging through advanced network architectures is illustrated in the left panel of Fig.~\ref{fig:money}. Relative to the ParticleNet and ParT baseline, pre-training on the JetClass dataset and fine-tuning leads to large gains for transformer architectures. An alternative path to enhance the performance is through Lorentz-equivariant networks. Lorentz-equivariant transformers allow us to combine these two improvements. In the right panel we show the training time required for some of the high-performance architectures. Here we observe a clear advantage of LLoCa compared to the original L-GATr architecture.

\subsection*{TopTagXL}

\begin{table}[t!]
    \centering
    \footnotesize
    \begin{tabular}{l lll}
        \toprule
        Network & Accuracy & AUC & $1/\epsilon_B$ ($\epsilon_S=0.8$) \\
        \midrule
        MIParT-L \cite{Wu:2024thh} & \result{0.9817}{\dz{}0.0001} & 0.9984 & \result{1795}{\dz{}41} \\
        LorentzNet* \cite{Gong:2022lye} & \result{0.9792}{\dz{}0.0001} & 0.9980 & \result{1014}{\dz{}24} \\ 
        L-GATr* \cite{Spinner:2024hjm} & \result{0.9818}{\dz{}0.0001} & 0.9985 & \result{1911}{\dz{}71} \\
\addlinespace[0.5ex]
        ParticleNet \cite{Qu:2019gqs} & \result{0.9792}{\dz{}0.0001} & 0.9980 & \dz{}\result{990}{\dz{}18} \\
        LLoCa-ParticleNet* & \result{0.9793}{\dz{}0.0001} & 0.9980 & \result{1046}{\dz{}23} \\
\addlinespace[0.5ex]
        ParT \cite{Qu:2022mxj} & \result{0.9817}{\dz{}0.0001} & 0.9984 & \result{1865}{\dz{}103} \\
        LLoCa-ParT* & \result{0.9818}{\dz{}0.0001} & 0.9985 & \result{1804}{\dz{}101} \\
\addlinespace[0.5ex]
        Transformer & \result{0.9809}{\dz{}0.0001} & 0.9983 & \result{1510}{\dz{}32} \\
        LLoCa-Transformer* & \result{0.9818}{\dz{}0.0001} & 0.9985 & \result{1842}{\dz{}56} \\
        \bottomrule
    \end{tabular}
    \caption{Top tagging accuracy, AUC and background rejection rates at fixed signal efficiency on the TopTagXL dataset. Lorentz-equivariant methods are indicated with an asterisk. Uncertainties are estimated using three random seeds.}
    \label{tab:topxl_tagging}
\end{table}

The main limitation in the established top tagging dataset discussed above is the small amount of training data, making careful hyperparameter tuning essential to avoid overfitting. While one million training events were sufficient at the time of publication~\cite{Kasieczka:2019dbj}, transformers require larger datasets. 

To fill this gap, we provide the large-scale TopTagXL dataset as an updated and enlarged version of the original top tagging dataset. 
It contains 135M events divided in 100M/25M/10M events for training, testing, and validation.
We use Pythia8.310~\cite{Bierlich:2022pfr} for the generation and the evolution to the final outgoing states. We include initial- and final-state radiation effects as well as multi-particle interactions. Pileup effects are excluded from this study.
The detector simulation uses the CMS detector as simulated by Delphes3.5.0~\cite{deFavereau:2013fsa}.
Compared to~\cite{Kasieczka:2019dbj}, we use more recent versions of Pythia8 and Delphes3, which include several improvements, and we include a larger set of features. 
The detector simulation uses the latest default CMS card, which contains a more accurate simulation of the detector calorimeters compared to the previous ATLAS detector. We also include a track smearing module based on the CMS detector.
We include the simulation cards for this dataset in our code repository.
The jet kinematic constraints for this dataset are
\begin{align}
p_{T,j} = 500,\ldots,1000\,\gev \qquad \text{and} \qquad |\eta_j| < 2.0 \;.
\end{align}
We store the four-momentum, PID, charge, and impact parameters of the leading 128 constituents, ordered by $p_T$.
For both top tagging datasets, jets are clustered using FastJet~\cite{Cacciari:2011ma} with the anti-$k_T$ algorithm and jet radius $R=0.8$. Only fully reconstructed top jets are used.

The resulting performance for our networks is reported in Table~\ref{tab:topxl_tagging}. We train all networks using the official implementation, and adopted the training setup used for ParT on the JetClass dataset~\cite{Qu:2022mxj}. We find that the Lorentz-equivariant transformers L-GATr, LLoCa-ParT and LLoCa-Transformer achieve the top performance, and are marginally better than the ParT network. On the other hand, the Lorentz-equivariant graph networks LorentzNet and LLoCa-ParticleNet are even worse than the non-equivariant transformer, ParT and MIParT-L. Once again, this highlights the necessity of using transformer architectures at scale.

\subsection*{Studying symmetry breaking}

\begin{table}[b!]
\centering
\begin{small}
\begin{tabular}{l ccccc}
\toprule
 & \multicolumn{2}{c}{Accuracy} &\;\;\;\;\;& \multicolumn{2}{c}{AUC} \\
Residual symmetry & Arch.&Input && Arch.&Input \\
\midrule
Non-equivariant & 0.855&0.864 && 0.9867&0.9882 \\
$\mathrm{SO}(2)_\text{beam}$ & 0.862&0.864 && 0.9878&0.9882 \\
$\mathrm{SO}^+(1,1)_\text{beam}\times \mathrm{SO}(2)_\text{beam}$ & 0.863&0.864&& 0.9881&0.9882\\
$\mathrm{SO}(3)$ & 0.859&0.860 && 0.9875&0.9876 \\
$\mathrm{SO}^+(1,3)$ (Lorentz) & \multicolumn{2}{c}{0.856} && \multicolumn{2}{c}{0.9870} \\
\bottomrule
\end{tabular}
\end{small}
\caption{Effect of symmetry-breaking for the LLoCa-Transformer on the JetClass dataset, using the LLoCa framework with either architecture-level ('Arch.') or input-level ('Input') symmetry breaking as discussed in Section~\ref{sec:symmetry_breaking}. The case ``$\mathrm{SO}(2)_\text{beam}$ + Input'' is our default LLoCa-Transformer jet tagger.}
\label{tab:jetclass_tagging_symmetrybreaking}
\end{table}

Finally, we perform ablation studies on the LLoCa-Transformer to analyze the impact of Lorentz symmetry breaking.
In Table~\ref{tab:jetclass_tagging_symmetrybreaking} we summarize the tagging performance on the JetClass dataset under different preserved symmetries. 
For jet tagging, we expect the tagging score to be invariant under the axial Lorentz subgroup $\mathrm{SO}^+(1,1)_\text{beam} \times \mathrm{SO}(2)_\text{beam}$, corresponding to boosts along and rotations around the beam axis. 
In our case, the non-equivariant baseline already respects this residual symmetry because the input features $E, p_T, \Delta R, \Delta \phi, \Delta \eta$ are themselves invariant under $\mathrm{SO}^+(1,1)_\text{beam} \times \mathrm{SO}(2)_\text{beam}$. Equivariant architectures then extend this baseline by introducing higher-order internal representations that transform non-trivially under the chosen symmetry group.

Full Lorentz equivariance without symmetry breaking yields only marginal improvements over the non-equivariant baseline, as the network predictions become overly constrained by the large symmetry group. Similarly, enforcing $\mathrm{SO}(3)$ equivariance leads to only minor gains.
In contrast, breaking symmetries at the architecture level, see Sec.~\ref{sec:symmetry_breaking}, provides substantial performance improvements. 
Preserving the residual symmetry $\mathrm{SO}^+(1,1)_\text{beam} \times \mathrm{SO}(2)_\text{beam}$ yields the best performance among the architecture-level symmetry-breaking configurations we tested. However, the choice of residual symmetry group is essential: breaking to a smaller group such as $\mathrm{SO}(2)_\text{beam}$ leads to a clear reduction in performance.

The best overall performance is achieved with the Lorentz-equivariant network in which symmetry breaking is applied at the input level, as described in Sec.~\ref{sec:symmetry_breaking}. In this case, the inductive bias of full Lorentz equivariance is retained, while the effective symmetry can be adjusted during training by increasing or decreasing the reliance on the symmetry-breaking inputs depending on the phase space region.
In fact, we observe that as long as the symmetry breaking inputs allow the network to reduce the equivariance to at least the symmetry realized in the data, $\mathrm{SO}(1,1)_\text{beam}\times \mathrm{SO}(2)_\text{beam}$, all options for input-level symmetry breaking achieve the same optimal performance -- even the setting where the residual symmetry group is empty.
In our jet tagging example, the configurations preserving $\mathrm{SO}^+(1,1)_\text{beam} \times \mathrm{SO}(2)_\text{beam}$ or only $\mathrm{SO}(2)_\text{beam}$, as well as the non-equivariant network with input-level symmetry breaking, all achieve the best AUC (0.9882) and best accuracy (0.864).

\section{Conclusions}
\label{sec:outlook}

Embedding symmetries in neural networks, especially Lorentz symmetries,
is a long-standing challenge in particle physics and modern ML. Current approaches 
are limited by computational cost, available geometric representations, 
and applicability to specialized architectures.
We have proposed Lorentz Local Canonicalization (LLoCa) as a flexible, high-performance framework which provides:
\begin{enumerate}
    \item exact Lorentz-equivariance at minimal computational costs for any network architecture;
    \item a straightforward way to break and even remove Lorentz-equivariance;
    \item efficient any-order input, output, and latent tensorial representations of the Lorentz group.
\end{enumerate}
We achieved this by using local canonicalization~\cite{lippmann2025beyond}
with tensorial message passing, extended to the Lorentz group.
LLoCa predicts a local reference frame for each particle in an equivariant manner.
Particle features in each local frame are Lorentz-invariant and can be transformed with any network architecture. We apply LLoCa to GNNs and transformers as the work horses of high-performance ML at the LHC.

We have illustrated the performance of LLoCa for amplitude regression, event generation, and jet tagging. We have obtained state-of-the-art performance for each task, starting from a non-equivariant network and elevating it to its Lorentz-equivariant counterpart. This step keeps the original network unchanged and increases the number of learnable parameters by approximately 1\%. 
LLoCa enables efficient Lorentz-equivariance; it reduces FLOPs by $10\times$ and training time by $4\times$ compared to the state-of-the-art L-GATr architecture while achieving the same performance.

Next, we have studied the effect of Lorentz-symmetry breaking for systems respecting only a residual symmetry group of the Lorentz group. For event generation, we have found that it is sufficient to have an architecture that is equivariant under the unbroken subgroup of the Lorentz group. For jet tagging, our studies have shown that including Lorentz-symmetry breaking information at the input level is needed for optimal performance. Moreover, once this information is provided, additional symmetry-breaking inputs do not degrade performance -- the network continues to achieve the same optimal accuracy.

Third, in our framework data augmentation appears as a byproduct of the definition of reference frames. It allows for a fair comparison of exact Lorentz-equivariant networks and data augmentation. 

In the jet tagging case, we have trained LLoCa versions of ParticleNet and ParT, established non-equivariant architectures. This demonstrates the flexibility of the  methodology and allows for a fair comparison between the standard and Lorentz-equivariant implementations. We have also performed an extended comparison on the novel TopTagXL~\cite{favaro_2025_10878355} dataset.

\subsection*{Data and code availability}
The LLoCa implementation is available at \url{https://github.com/heidelberg-hepml/lloca}, and all experiment code at \url{https://github.com/heidelberg-hepml/lloca-experiments}.
The amplitude regression datasets are available at \href{https://doi.org/10.5281/zenodo.16793011}{Zenodo.16793011} while the TopTagXL dataset is available at \href{https://doi.org/10.5281/zenodo.10878355}{Zenodo.10878355}.

\section*{Acknowledgements}
We thank Víctor Bresó for valuable discussions and generating the amplitude regression datasets. We would also like to thank Gregor Kasieczka for assistance when creating the TopTagXL dataset, and Vishal Ngairangbam for helpful discussions on symmetry breaking.

L.F. is supported by the Fonds de la Recherche Scientifique - FNRS under Grant No. 4.4503.16.
G.G. is supported by the Deutsche Forschungsgemeinschaft (DFG, German Research Foundation) under grant 554514035.
J.S. and T.P. are supported by the Deutsche Forschungsgemeinschaft (DFG, German Research Foundation) under grant 396021762 – TRR 257 Particle Physics Phenomenology after the Higgs Discovery. J.S. is funded by the Carl-Zeiss-Stiftung through the project Model-Based AI: Physical Models and Deep Learning for Imaging and Cancer Treatment. P.L. and F.A.H. acknowledge funding by Deutsche Forschungsgemeinschft (DFG, German Research Foundation) – Projektnummer 240245660 - SFB 1129.
This work is supported by Deutsche Forschungsgemeinschaft (DFG) under Germany’s Excellence Strategy EXC-2181/1 - 390900948 (the Heidelberg STRUCTURES Excellence Cluster).

Computational resources have been provided by the supercomputing facilities of the Université catholique de Louvain (CISM/UCL) and the Consortium des Équipements de Calcul Intensif en Fédération Wallonie Bruxelles (CÉCI) funded by the Fond de la Recherche Scientifique de Belgique (F.R.S.-FNRS) under convention 2.5020.11 and by the Walloon Region.
The present research benefited from computational resources made available on Lucia, the Tier-1 supercomputer of the Walloon Region, infrastructure funded by the Walloon Region under the grant agreement n°1910247.
The authors acknowledge support by the state of Baden-Württemberg through bwHPC and the German Research Foundation (DFG) through the grants INST 35/1597-1 FUGG and INST 39/1232-1 FUGG.
The authors gratefully acknowledge the computing time provided on the high-performance computer HoreKa by the National High-Performance Computing Center at KIT (NHR@KIT). This center is jointly supported by the Federal Ministry of Education and Research and the Ministry of Science, Research and the Arts of Baden-Württemberg, as part of the National High-Performance Computing (NHR) joint funding program (https://www.nhr-verein.de/en/our-partners). HoreKa is partly funded by the German Research Foundation (DFG).

\appendix

\section{Training and network specifications}
\label{app:training}

\subsection*{Frames-Net}
\label{app:framesnet}

We use the same architecture for the Frames-Net, $\varphi$ in Eq.~\eqref{eq:construct_v}, for all LLoCa networks. It consists of 2 layers with 128 hidden channels and GELU activations. In Figure~\ref{fig:ampxl_abl_method} we show as ``F-Net small'' a Frames-Net with only 16 hidden channels, and as ``F-Net dropout'' a Frames-Net with 128 hidden channels and a dropout rate of 0.2. We perform an additional normalization step on the results of Eq.~\eqref{eq:construct_v}
\begin{equation}
    v_{i,k} \gets v_{i,k} \Big/ \sqrt{\sum_{j=1}^N \| v_{i,k}\|^2}.
\end{equation}

In general, we take extra care to avoid numerical instabilities in the 
orthonormalization procedure described in Eq.~\eqref{eq:gram-schmidt}.
Before any training, we introduce a small mass regulator $m_\epsilon$
to ensure that all particles have positive masses. This is done by increasing the 
energy as $E'=\sqrt{m_\epsilon^2 + E^2}$. In our applications, we use $m_\epsilon=10^{-5}$ for amplitude regression and $m_\epsilon=5\cdot10^{-3}$ for
jet tagging.

Then we verify that $v_0$ is not lightlike, as this would lead to an ill-defined 
boost $B$. We check that $|v_0| < \epsilon_\text{lightlike}$ with $\epsilon_\text{lightlike}=10^{-16}$. If this condition is not satisfied, 
we replace $v_0$ with $v_0 + \epsilon_\text{lightlike}\delta$, with $\delta$ 
a timelike noise vector constructed by sampling the components of a vector from
a Gaussian distribution, taking the absolute value, and setting the energy
component to two times the spatial norm. 
Finally, the $\mathrm{SO}(3)$ Gram-Schmidt orthogonalization used to construct the
rotation matrix $R$ requires linearly independent vectors. We numerically control
this condition by verifying that $\|\vec w_1 \times \vec w_2\| < \epsilon_\text{collinear}$.
If a regularization is needed, we replace both vectors with $\vec w_1 + \epsilon_\text{collinear}\vec \delta_1$ and $\vec w_2 + \epsilon_\text{collinear}\vec \delta_2$, where $\vec \delta_{1,2}$ are random normal directions, i.e.~$\delta_i \sim\mathcal{N}(0,1)$. 
This choice of regularization allows for safe prediction of local frames also in
the edge case of $N<3$ input particles. For both $N=1$ and $N=2$, the cross product 
between $\vec w_1$ and $\vec w_2$ would be zero; the regularization ensures that
the vectors are linearly independent but at the cost of Lorentz-equivariance
violation. 
In our studies, we use $\epsilon_\text{collinear}=10^{-16}$, and we do not observe regularization throughout trainings besides a few
occurrences at initialization, validating the numerical stability of our orthogonalization method.

\subsection*{Amplitude regression}
To stabilize the network performance, we boost the inputs in the reference frame of the sum of the two incoming particles, also called center-of-mass frame. Then, we apply a random general Lorentz transformation.

The GNN used for the amplitude regression results uses standardized four-momenta and one-hot encoded particle types as inputs. 
The GNN is a standard fully connected
graph neural network with edge convolutions. The message passing operation
uses the hidden node representations, and the Minkowski product as an additional edge feature.
The neural network is an MLP with three hidden layers with 128 hidden channels
each, and the messages are aggregated with a summation.
The large-scale architecture uses three edge convolution blocks.
The LLoCa-GNN uses the same architecture and we use 64 scalars and 16 vectors
as internal representation. 

The transformer uses the same particle features as the GNN as inputs, which are embedded
into a latent space with 128 channels using a linear layer.
Then, we apply two residual transformations to the inputs. First, 
we apply a standard multi-head self-attention block with eight heads. Then,
we transform the inputs with a fully-connected network with two layers
and GELU nonlinearities.
The large-scale architecture includes eight of these blocks.
The LLoCa-Transformer has the same architecture. We specify the internal
representation of a single head to be eight scalars and two vectors,
i.e.~a 50\% split between scalar and vector channels.  

The L-GATr network is taken from the official repository but with 20 hidden multivector channels, while the MLP-I
is a standard MLP with all the Mandelstam variables as input. The architecture
consists of five layers with 128 hidden channels each and GELU nonlinearities.
With these settings, the GNN and the LLoCa-GNN have $2.5 \cdot 10^{5}$ and 
$2.7 \cdot 10^{5}$ parameters each; the transformer, LLoCa-Transformer, and
L-GATr have $\sim 10^{6}$ parameters; the MLP-I totals $5.5 \cdot 10^{4}$ learnable parameters.  

Data augmentation is implemented through the same polar decomposition theorem
used for Eq.~\eqref{eq:polar_decomposition}. We sample a rotation matrix
from a uniform distribution on a 3D-sphere. The non-compactness of the
Lorentz group does not allow to uniformly sample boosts. Therefore, we sample
the components of $\vec \beta$ from $\mathcal{N}(0.0, 0.1)$. The full
Lorentz-transformation is applied after boosting the particles to the 
center-of-mass frame of the two incoming particles.
We find that this setup provides the best performance from a limited 
hyperparameters scan. 

All the networks are trained on a single A100 GPU, using the Adam optimizer
with \\$\beta = [0.99, 0.999]$, a batch size of 1024, and for $2\cdot 10^{5}$
iterations. We use the \\\texttt{ReduceLROnPlateau} learning rate scheduler
which reduced the learning rate by 0.3 if no improvement has been observed
in the last 20 validations. The networks are implemented in PyTorch.
The timing information is extracted from the full training of a neural network on a H100,
while the FLOPs are counted with \texttt{torch.utils.flop\_counter.FlopCounterMode} from a single forward pass.

\subsection*{Event generation}

Our transformer and L-GATr generators follow exactly the same setup as in Ref.~\cite{Brehmer:2024yqw}. The transformer uses 108 channels, 8 heads and 6 layers with GELU activations, totally $5.7\cdot 10^5$ parameters. The L-GATr network uses 32 scalar channels, 16 multivector channels, 8 heads, 6 blocks and GELU activations, totalling $5.4\cdot 10^5$ parameters. The LLoCa-Transformer uses the same backbone architecture as the transformer, and the same Frames-Net architecture as described in Appendix~\ref{app:framesnet}. The DA-Transformer also uses the same architecture as the transformer, but augments events during training with the same random Lorentz transformations as described above for amplitude regression. 

All generators are trained using the same hyperparameters. For each multiplicity, we split the dataset into 98\% for training and 1\% each for validation and testing. We use the Adam optimizer with learning rate $10^{-3}$ and default settings otherwise. The learning rate is reduced by a factor of 10 if the validation loss shows no improvement for 20 consecutive validation steps. We train for 200000 iterations using a batch size of 2048 and validate every 1000 steps.

The neural classifier used in Appendix~\ref{app:generation-extra} also follows the setup of Ref.~\cite{Brehmer:2024yqw}. We use a simple MLP consisting of 3 layers with 256 channels each. It is trained on a binary cross-entropy objective using 1M generated events and the full truth dataset, with an 80\%/10\%/10\% split for training, testing, and validation, respectively. The classifier inputs include the complete events in the $x$ representation defined in Eq.~\eqref{eq:momentum_rep}, augmented by all pairwise $\Delta R$ features, as well as the $x$ representation of the reconstructed particles $t, \bar t, W^+, W^-$. Training is conducted over 500 epochs with a batch size of 1024, a dropout rate of 0.1, and the Adam optimizer with default parameters. We start with a learning rate of 0.0003 and reduce it by a factor of 10 if the validation loss shows no improvement for 5 consecutive epochs.

\subsection*{Jet tagging}

We use the official implementation for all baseline jet taggers, namely ParT~\cite{Qu:2022mxj}, ParticleNet~\cite{Qu:2019gqs,Qu:2022mxj}, LorentzNet~\cite{Gong:2022lye}, MIParT~\cite{Wu:2024thh}, PFN~\cite{Qu:2022mxj}, P-CNN~\cite{Qu:2022mxj} for our studies. For each network, we have confirmed that we reproduce the results reported in the official implementations. The parameter counts of all networks are given in Table~\ref{tab:jetclass_tagging}.
Our baseline transformer architecture is designed to follow the ParT architecture, but without the learnable attention bias and with mean aggregation instead of class attention aggregation. It closely follows the ParT architecture size, using ten attention blocks, 128 hidden channels, 8 attention heads and expands the hidden units in each attention block's MLP by a factor of four. 

To implement tensorial message passing following LLoCa, we modify the official ParT and ParticleNet implementations as follows. For LLoCa-ParticleNet, we transform the sender's local features from its own local frame to that of the receiver, and perform the k-nearest-neighbors search using four-momenta expressed in each node's local frame. In LLoCa-ParT, we similarly evaluate the edge features, which are used to compute the attention bias, based on four-momenta in local frames. We implement tensorial message passing in the particle attention blocks, but keep standard scalar message passing for class attention. We find that the performance of LLoCa-ParT does not change when using automatic mixed precision in the backbone, however we ran the timing, memory consumption and FLOPs benchmarks in single precision to simplify the comparison with other architectures.

We evaluated the timings, memory consumption and FLOPs reported in Table~\ref{tab:jetclass_tagging} for all networks in our own environment. The timings are obtained from training for 1000 iterations on a H100 GPU without validation and with a fraction of the dataset stored in memory and then extrapolating the result to 1M iterations. The GPU memory usage is evaluated using the command \texttt{torch.cuda.max\_memory\_allocated()} in the same environment as used for the timing measurement. The FLOPs per forward pass are computed for events containing 50 particles using PyTorch's \texttt{torch.utils.flop\_counter.FlopCounterMode} utility. 

We emphasize that timing and memory usage depend on the implementation. The metrics reported in Table~\ref{tab:jetclass_tagging} should be understood as an upper bound, with room for improvements in the future. A key aspect is the zero-padding of jets to a fixed maximum number of particles, 128, as done in ParT, ParticleNet, MIParT, PFN and P-CNN. While convenient to implement, the zero-padded jet consistuents have to be saved and operated on throughout the architecture, which significantly increases the memory usage and timing, respectively. Instead, LorentzNet is implemented as a  \texttt{torch\_geometric.nn.MessagePassing} module, which natively supports operations on graphs with variable number of nodes. We have slightly adapted the official LorentzNet implementation to make use of this feature. Our transformer and L-GATr networks both use the \texttt{xformers.ops.memory\_efficient\_attention} backend, which supports graphs with a variable number of nodes via the \\\texttt{xformers.ops.fmha.BlockDiagonalMask} module.

\subsection*{JetClass and TopTagXL}
For the JetClass studies, all networks are trained on the official JetClass dataset~\cite{Qu:2022mxj}, which includes 100M events for training, 20M for testing and 5M for validation. Accuracy and AUC are used as global evaluation metrics following Ref~\cite{Qu:2022mxj}, and class-specific background rejection rates at fixed signal efficiency are shown in Appendix~\ref{app:tagging-extra}. 
We adopt the official training procedure for PFN, P-CNN, ParT and L-GATr. Our transformer uses the same setup as ParT. The LLoCa version is always trained with the same training hyperparameters as its non-equivariant counterpart. For ParticleNet, we find that the official ParT training setup yields significantly better results than the official ParticleNet setup, especially for the LLoCa versions. We therefore use the ParT training procedure also for ParticleNet. LorentzNet, PELICAN-lite and MIParT-L are also trained with the ParT training procedure.

The TopTagXL dataset consists of 100M events for training, 25M for testing, and 10M for validation. All networks are trained for 1M iterations using the AdamW optimizer with batch size 512, initial learning rate $10^{-3}$ and a cosine annealing learning rate schedule.

\subsection*{Top tagging}
For the top tagging results in Tab.~\ref{tab:top_tagging}, we take the results
of previous networks from the literature and we only focus on the training
of the LLoCa-Transformer. The network architecture and the training parameters
are tailored to this specific dataset to deal with its limited statistics.
The LLoCa-Transformer has a 12 blocks for a total of $1.6\cdot 10^{6}$ 
parameters. We use eight heads in the self-attention operation with 12
scalars and a single vector for each head. In the following MLP, we expand
the latent features by a factor two instead of four.
The optimization is done with the Lion~\cite{chen2023symbolicdiscoveryoptimizationalgorithms} optimizer with
$\beta=[0.9, 0.999]$ and a \texttt{CosineAnnealingWarmRestarts} learning rate
scheduler. We train for 300000 iterations, starting a new cycle once after
150000 iterations, with batch size 128 and initial learning rate of $3\cdot 10^{-5}$.
To regularize the training, we use a weight decay equal
to two in the main network and a dropout of 0.1 in the Frames-Net.
We observe improved training dynamic and better results if, prior to training,
a global boost to the restframe of the jet is applied to all the constituents.

\section{Training stability and equivariance violation}
\label{app:tests}
In this section, we test the stability of LLoCa in predicting local frames $L$ and the violation of equivariance, or invariance. As a representative sample, we extract this information from 100k events of the test dataset. \medskip
\begin{figure}[tb]
    \centering
    \includegraphics[width=0.323\linewidth]{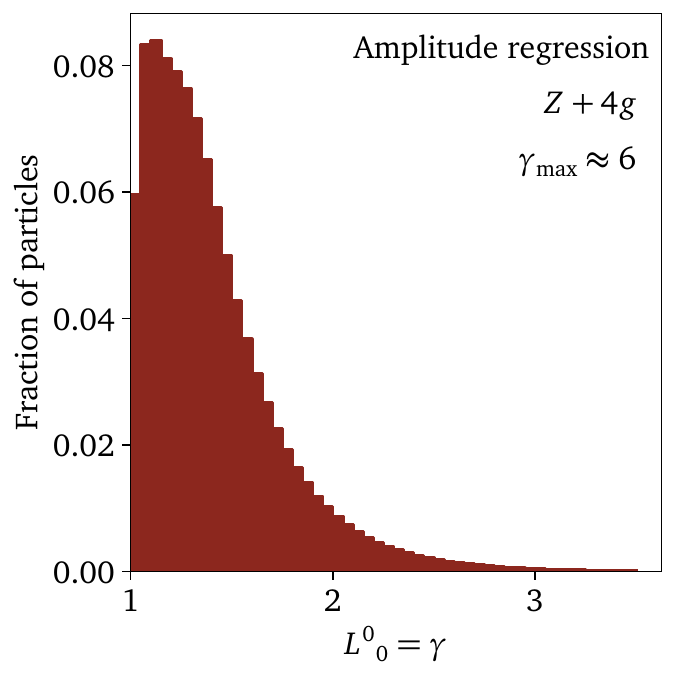}
    \includegraphics[width=0.323\linewidth]{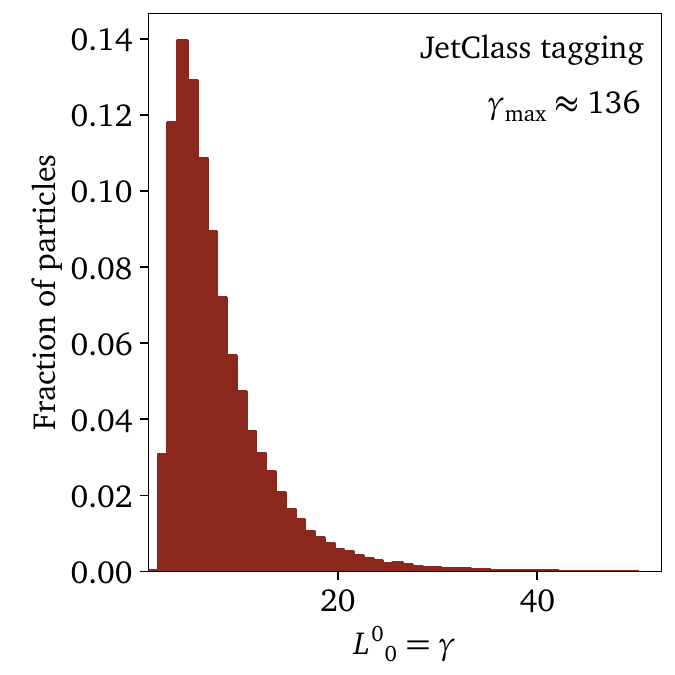}
    \includegraphics[width=0.323\linewidth]{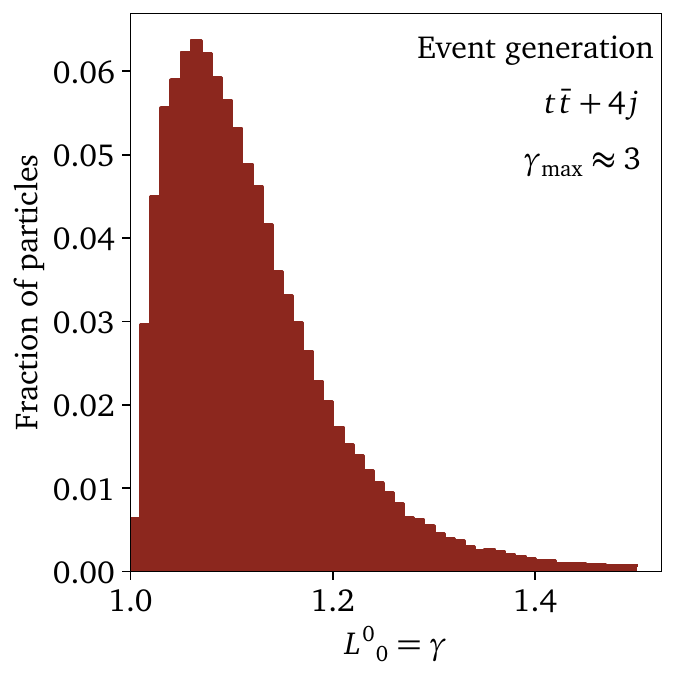}
    \caption{Distribution of local frame boost factors $\gamma$ of the local frames. From left to right: amplitude regression, jet tagging, and event generation.}
    \label{fig:app_gamma}
\end{figure}
\begin{figure}[tb]
    \centering
    \includegraphics[width=0.323\linewidth]{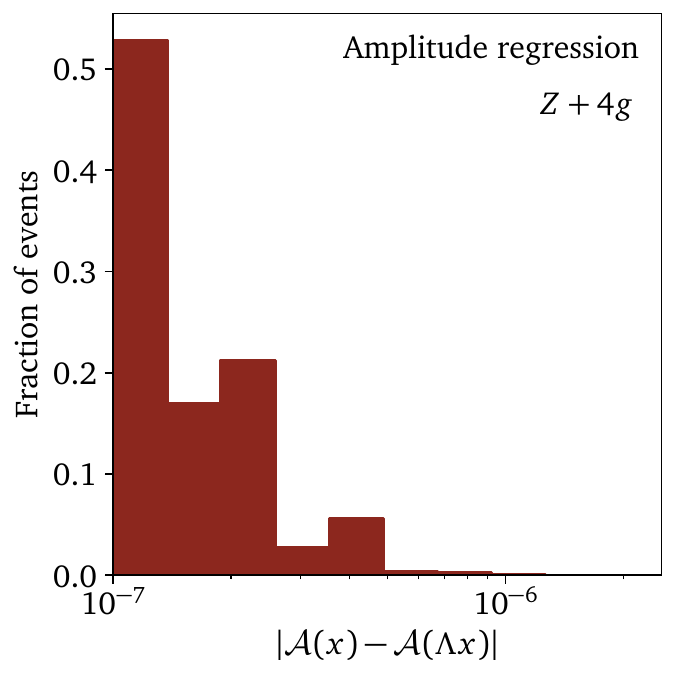}
    \includegraphics[width=0.323\linewidth]{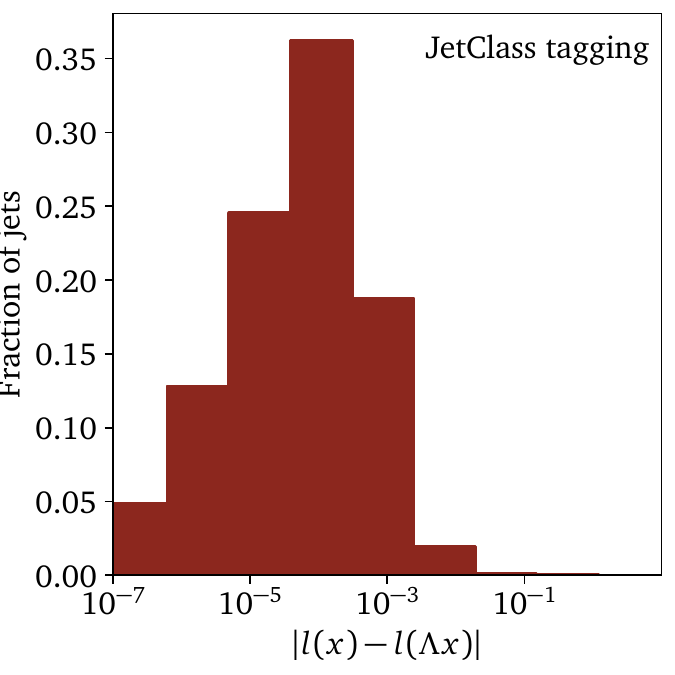}
    \includegraphics[width=0.323\linewidth]{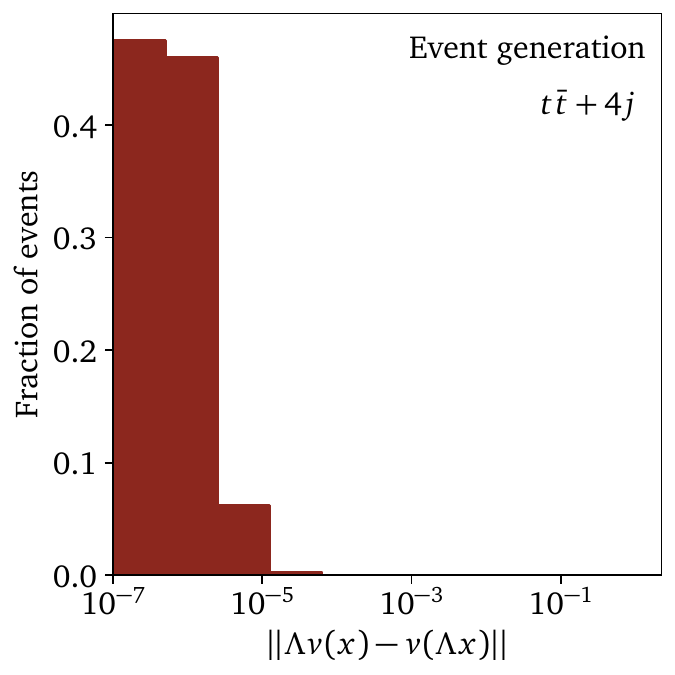}
    \caption{LLoCa network output invariance tests for amplitude regression (left), jet tagging (middle), and equivariance test for event generation (right), using random transformations $\Lambda$ as described in the text.}
    \label{fig:app_equi}
\end{figure}
We start by showing the distribution of the entry $L{^0}_0$ of the local frames, which corresponds to the boost factor $\gamma$ of the Lorentz transformation.
Figure~\ref{fig:app_gamma} shows the $\gamma$-factors for amplitude regression, jet tagging, and event generation. In all cases, the distribution is smooth with a tail extending towards larger boosts. The JetClass dataset contains boosted jets and this property is reflected in the overall larger boosts in the predicted local frames. We indicate with $\gamma_\mathrm{max}$ the largest boost in the test set. The boosted jet tagging study shows the largest boost factor, $\gamma_\mathrm{max}=136$, as expected from the kinematic range of the jets. \medskip

\begin{figure}
    \centering
    \includegraphics[width=0.4\linewidth]{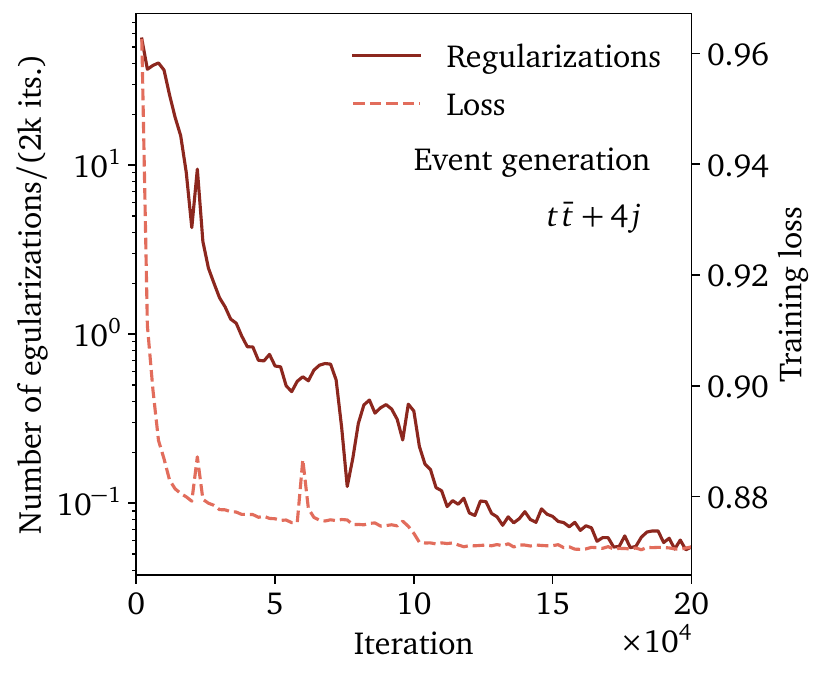}
    \caption{Average number of regularizations every 2000 iterations over the training process for the LLoCa event generation. The regularization ensures that $\gamma <3$.}
    \label{fig:app_evgen_regs}
\end{figure}

Figure~\ref{fig:app_equi} shows the equivariance, or invariance, violation from numerical effects. Our test analyzes the variation in the network output after applying a Lorentz transformation which respects the symmetry of the application. The Lorentz transformation is a general boost for the amplitude regression study, a $\mathrm{SO}^+(1,1)_\text{beam}\times \mathrm{SO}(2)_\text{beam}$ transformation for jet tagging, and a $\mathrm{SO}(2)_\text{beam}$ transformation for event generation. We sample the Lorentz transformation from the polar decomposition $\Lambda=RB$, i.e. a combination of a pure boost and a rotation. We sample the rotations uniformly however, since the Lorentz group is non-compact, we limit the range of possible boost velocities in the three space directions to $\beta^1,\beta^2,\beta^3\sim\mathcal{N}(0, 0.1)$. 
LLoCa networks show a remarkably small equivariance violation for amplitude regression and event generation. In these two cases, the observed deviations are compatible with the single-float numerical precision used in the implementation. For JetClass, LLoCa shows a longer tail and a mean invariance violation of $10^{-4}$. We noted that this behavior is connected to frame-to-frame transformations with large boost factors, and we leave the improvement of this aspect of LLoCa for future work. \medskip
Finally, we study in more detail the regularization introduced in Sec~\ref{sec:generation}. To achieve a stable prediction of the vector field used for event generation, we introduce a soft-clamping of the $\gamma$-factors. The implementation applies the transformation
\begin{align}
    \gamma_\mathrm{reg} &= \gamma_\mathrm{th} - \mathrm{Softplus}(\gamma_\mathrm{th} - \gamma; s)
\end{align}
to the predicted boost $\gamma$, where $\gamma_\mathrm{th}$ indicates the maximum reachable boost. The optimal threshold value depends on the dataset, and it should be considered a training hyperparameter in cases where LLoCa shows training instabilities. For event generation, we set $\gamma_\mathrm{th}=3$. The $\mathrm{Softplus}$ operation is defined as 
\begin{equation}
    \mathrm{Softplus}(x;s) = \frac{1}{s}\log (1+e^{sx})  \,, 
\end{equation}
where the parameter $s$ controls the rate at which the threshold value is reached, with larger $s$ corresponding to larger curvatures. Overall, the regularization keeps the boost factors far from the threshold unchanged, while the others are smoothly regularized.
Figure~\ref{fig:app_evgen_regs} shows, for the $t\bar{t}+4j$ event generation study, the average number of regularizations applied to the local frames as a function of the training iterations. We include the average loss value in the same figure. At initialization, we observe a large number of regularizations which decreases as the loss converges. During the final iterations, the regularization is activated on average $<1$ times per batch, indicating that it allows stable initial training without altering the converged network. 

\section{PELICAN-lite}
\label{app:pelican}

This section provides a brief introduction to the PELICAN architecture and outlines the technical differences between our PELICAN-lite implementation\footnote{\url{https://github.com/heidelberg-hepml/pelican}} and the official PELICAN~\cite{Bogatskiy:2022czk,Bogatskiy:2023nnw} implementation\footnote{\url{https://github.com/abogatskiy/PELICAN}}.
\medskip

PELICAN (Permutation-Equivariant and Lorentz Invariant or Covariant Aggregator Network) achieves Lorentz invariance by transforming four-momenta into Lorentz-invariant features using Minkowski inner products, which serve as pairwise or edge features. A central concept in this framework is the use of representations of the permutation group, such as graph-level (rank 0), node-level (rank 1), and edge-level (rank 2) representations. PELICAN embeds edge-level inputs -- and optionally additional graph-level or node-level inputs -- into edge-level feature spaces, processes them using generic permutation-equivariant operations, and subsequently projects them onto graph-level, node-level, or edge-level outputs.

The core building blocks of PELICAN are the $\mathrm{Eq}_{n\to m}$ layers, which are learnable projections from rank-$n$ to rank-$m$ representations of the permutation group. Each $\mathrm{Eq}_{n\to m}$ layer combines a deterministic aggregation operation $\mathrm{Agg}_{n\to m}$ with the standard components of a feed-forward network: linear layers, activation functions $\mathrm{Act}$, dropout, and layer normalization $\mathrm{LayerNorm}$. Assembling these components, the PELICAN-lite network with output rank $n$, denoted $\mathrm{PELICAN}^n$, can be expressed as
\begin{align}\label{eq:pelican}
    \mathrm{Agg}_{n\to m}(x) &= \text{Concatenate}\left( \left\{ \alpha_k \mathrm{AggMap}_k(x) \right\}_{k=1}^{B_{n+m}}\right) \notag\\
    \mathrm{Eq}_{n\to m} (x) &= \mathrm{Act} \circ \mathrm{Linear} \circ \mathrm{Agg}_{n\to m} \circ \mathrm{Dropout} \circ \mathrm{LayerNorm} \circ \mathrm{Act} \circ \mathrm{Linear}\;, \notag \\
    \mathrm{Embedding}(g, s_i, e_{ij}) &= \mathrm{Eq}_{2\to 2}\left(\mathrm{Concatenate}\left(\left\{\mathrm{Eq}_{0\to 2}(g), \mathrm{Eq}_{1\to 2}(s_i), e_{ij})\right\}\right)\right)\;, \notag\\
    \mathrm{PELICAN}^n (g, s_i, e_{ij}) &= \mathrm{Eq}_{2\to n} \circ \mathrm{Eq}_{2\to 2}\circ \cdots \circ \mathrm{Eq}_{2\to 2} \circ \mathrm{Embedding}(g, s_i, e_{ij})\;.
\end{align}

The operation $\mathrm{Agg}_{n\to m}$ constructs all possible aggregation maps $\mathrm{AggMap}_k$ corresponding to rank-$n$ to rank-$m$ mappings, applied independently to each channel. This results in a higher-dimensional representation of the latent space. The number of such aggregations in $\mathrm{Agg}_{n\to m}$ is given by the Bell number $B_{n+m}$, where $B_0=B_1=1$, $B_2=2$, $B_3=5$, and $B_4=15$. Explicit expressions for the aggregation maps can be found in Ref.~\cite{pmlr-v151-pan22a}.
To compensate for the $B_{n+m}$-fold increase in the number of channels, the first linear layer in the definition of $\mathrm{Eq}_{n\to m}$ typically projects to a lower-dimensional latent space. This space is then expanded by the $B_{n+m}$ aggregation maps, after which the second linear layer reduces it back to the original latent dimensionality. We find that introducing a learnable coefficient $\alpha_k$ for each aggregation map further enhances the expressivity of the architecture.

The architecture defined in Eq.~\eqref{eq:pelican} differs from the original PELICAN implementation in two aspects. First, we employ layer normalization instead of batch normalization. Batch normalization is generally discouraged in this context, as it exchanges information between different batch elements, making both training and evaluation dependent on the batch size. Second, the original implementation defines the aggregation map coefficient as $(N / N_\mathrm{mean})^{\alpha_k}$, where $N$ denotes the number of particles in the event and $N_\mathrm{mean}$ is the average number of particles across events. In contrast, we use the simpler form $\alpha_k$.

Our implementation reduces memory consumption by representing events of variable length $N$ with sparse rather than zero-padded dense tensors. A dense batch of $B$ events has shape $(B, N_\mathrm{max}, C)$, while the sparse form uses $(N_\mathrm{tot}, C)$ with $N_\mathrm{tot} = B N_\mathrm{mean}$. The dense representation introduces an overhead factor
\begin{equation}
g = \frac{B N_\mathrm{max}}{N_\mathrm{tot}} = \frac{N_\mathrm{max}}{N_\mathrm{mean}} \ge 1 \; .
\end{equation}
which grows with batch size (e.g.~$g \sim 2.5$ for $B=100$ on the top-tagging dataset). Since PELICAN operates on edge features, the resulting memory overhead scales as $g^2$, giving a reduction by roughly a factor of $g^2 \sim 8$ when using sparse tensors. Additionally, using \texttt{torch.compile} to fuse aggregation-map construction yields a $3\times$ speedup for the 208k-parameter network on an H100 GPU.

The PELICAN-lite performance and computational cost reported in Tab.~\ref{tab:jetclass_tagging} do not use the official PELICAN configuration with 132 hidden channels, 78 aggregation-map channels, and factorized weights. Instead, we use a reduced configuration with 72 hidden channels, 42 aggregation-map channels, and no factorized weights. In our implementation, the official configuration contains 172k parameters, while our “fair comparison” variant has 244k parameters; both are comparable in size to the LorentzNet and ParticleNet baselines. We do not train a PELICAN model with factorized weights because this architecture has the same memory and runtime cost as the corresponding non-factorized architecture with 816k parameters, making PELICAN approximately a factor of two slower and more memory-intensive than LorentzNet and ParticleNet.

\section{Additional results}

\subsection*{Event generation}
\label{app:generation-extra}

To further study the quality of the different event generators discussed in Section~\ref{sec:generation}, we train neural classifiers as an additional performance metric~\cite{Das:2023ktd, Grossi:2025pmm}. The classifiers are trained to distinguish the generated events from the ground truth, see App.~\ref{app:training} for details on the training procedure.

\begin{figure}
    \includegraphics[width=0.495\textwidth]{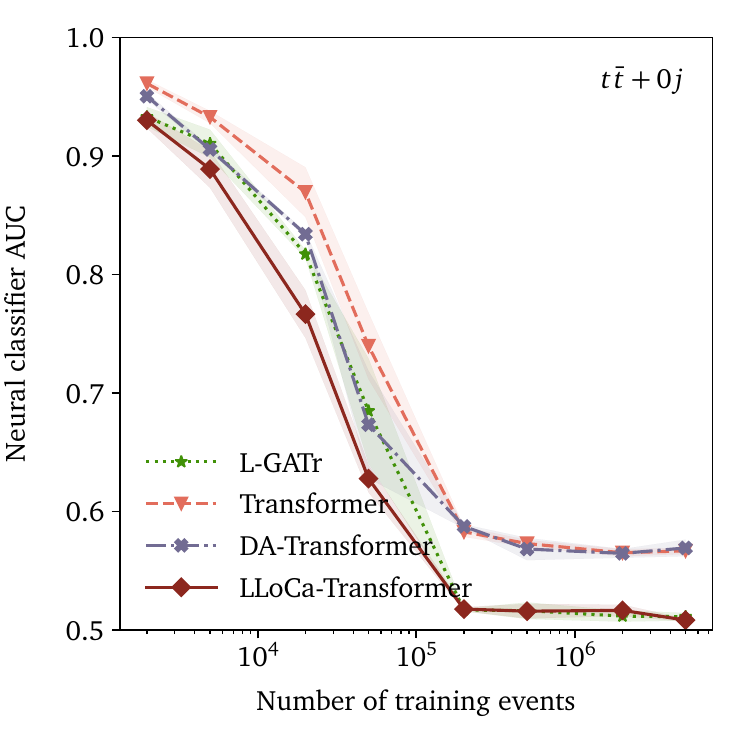}
    \includegraphics[width=0.495\textwidth]{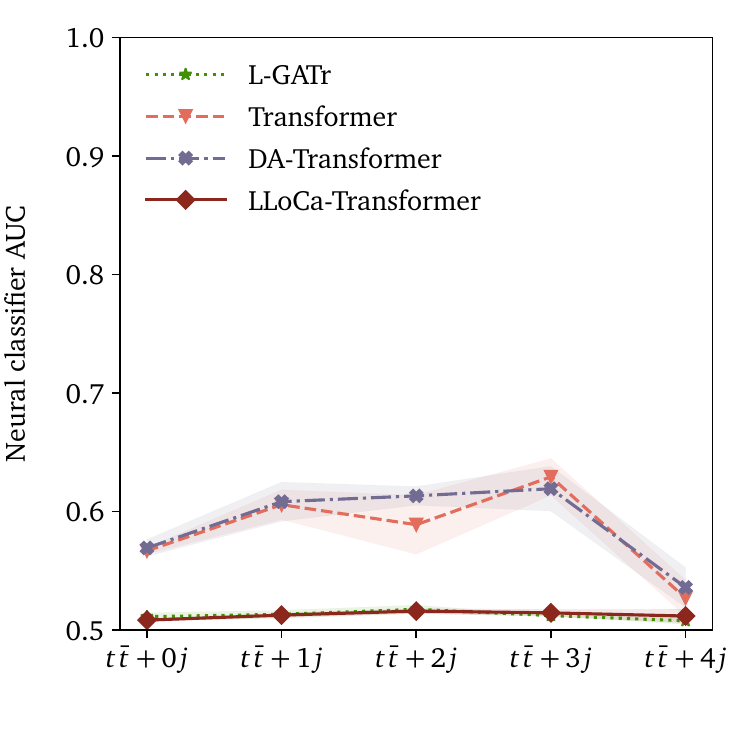} \\
    \includegraphics[width=0.495\textwidth]{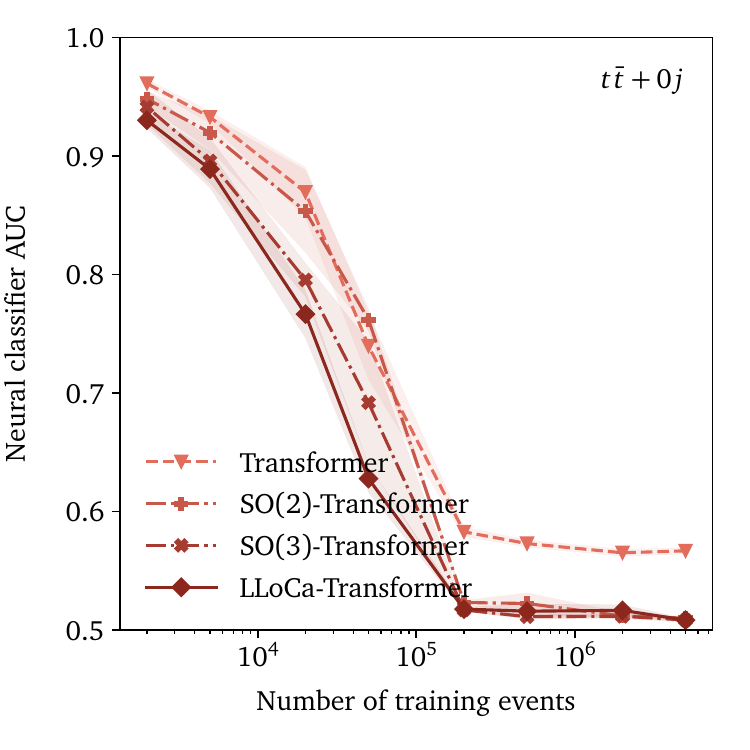}
    \includegraphics[width=0.495\textwidth]{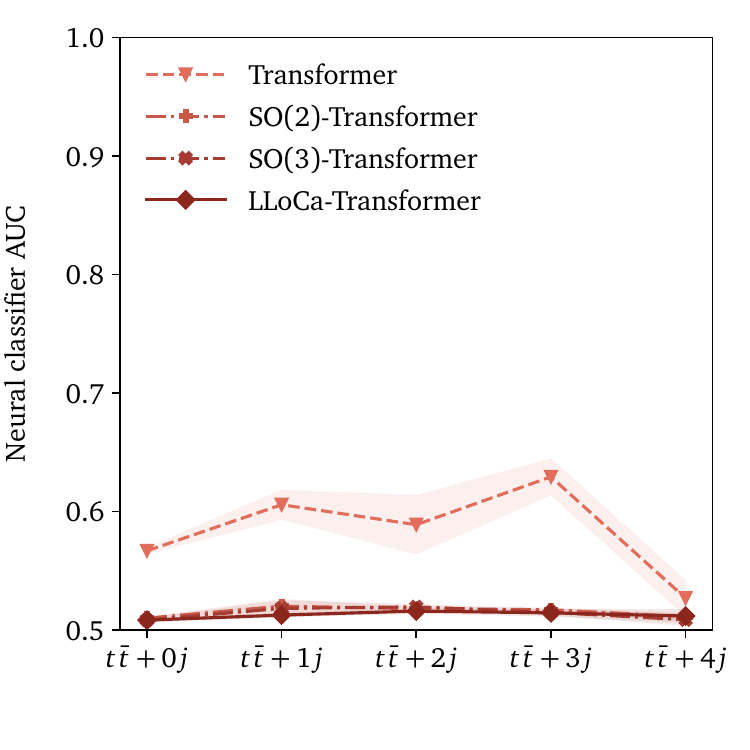}
    \caption{Upper panels: Classifier AUC of the LLoCa-Transformer compared with a non-equivariant transformer, a transformer trained with data augmentation, and L-GATr for event generation. Lower panels: Impact of explicit symmetry breaking on event generation, measured by classifier AUC. We compare a fully Lorentz-equivariant transformer with $\mathrm{SO}(2)$- and $\mathrm{SO}(3)$-equivariant transformers.}
    \label{fig:eventgen-scaling2}
\end{figure}

As shown in Figure~\ref{fig:eventgen-scaling2}, the neural classifier can not distinguish samples from the L-GATr and LLoCa-Transformer as well as the SO(2)- and SO(3)-Transformer generators if sufficient training data is used, resulting in an AUC of 0.5. The classifier can always distinguish events generated by the transformer and DA-Transformer generators from the ground truth, and both approaches show similar performance overall.
In the small-data regime, differences between the networks are overshed by uncertainties from the training process. 
Overall, the classifier metric supports the results found with the negative log-likelihood metric discussed in Section~\ref{sec:generation}. 

\subsection*{Jet tagging}
\label{app:tagging-extra}

\begin{figure}[t!]
    \centering
    \includegraphics[width=0.495\linewidth,page=1]{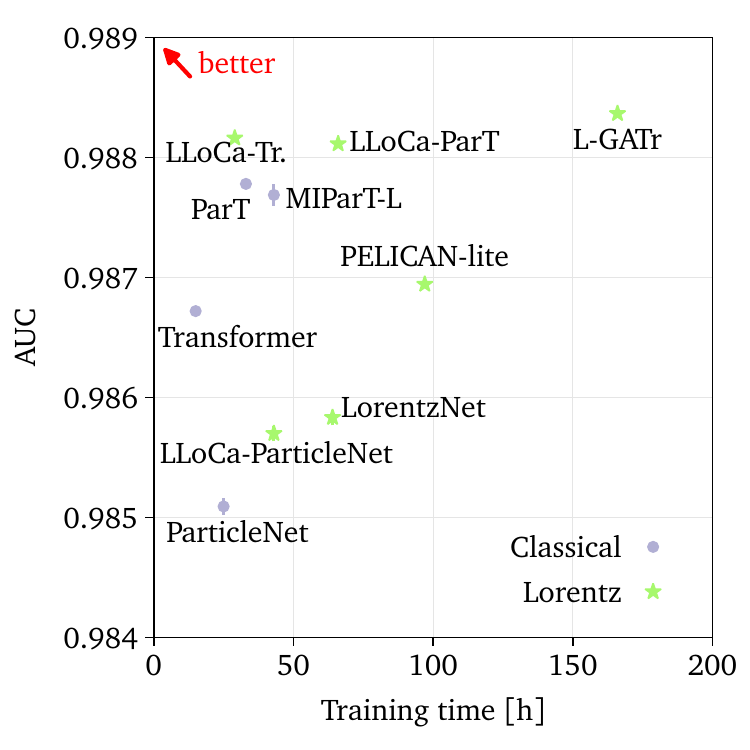}
    \includegraphics[width=0.495\linewidth,page=2]{figs/jetclass_landscape.pdf}
    \includegraphics[width=0.495\linewidth,page=3]{figs/jetclass_landscape.pdf}
    \includegraphics[width=0.495\linewidth,page=4]{figs/jetclass_landscape.pdf}
    \caption{Performance vs training time (top left), FLOPs (top right), memory usage (bottom left) and parameter count (bottom right) for top-performing jet taggers on the JetClass dataset. These figures are a visual representation of the results presented in Table~\ref{tab:jetclass_tagging}.}
    \label{fig:jetclass_landscape2}
\end{figure}

In Figure~\ref{fig:jetclass_landscape2} we provide visual representations of the results presented in Table~\ref{tab:jetclass_tagging}.

\bibliographystyle{SciPost-bibstyle}
\bibliography{literature}

\end{document}

%% file: incl_settings.tex
\usepackage[utf8]{inputenc} 
\usepackage[T1]{fontenc} 	
\usepackage[english]{babel} 

\usepackage[bitstream-charter]{mathdesign}
\usepackage{geometry} 		
\usepackage{amsmath} 		
\usepackage{amsthm} 		
\usepackage{mathtools} 		
\usepackage{float} 			
\usepackage{graphicx} 		
\usepackage{tabularx} 		
\usepackage{booktabs} 		
\usepackage{color, xcolor} 	
\usepackage{pdfpages} 		
\usepackage{extarrows} 		
\usepackage{multirow} 		
\usepackage{multicol} 		
\usepackage{caption} 		
\usepackage{subcaption} 	
\usepackage{enumitem} 		
\usepackage{setspace} 		
\usepackage{xspace} 		
\usepackage{ragged2e} 		
\usepackage{stackrel} 		
\usepackage{tikz} 			
\usetikzlibrary{shapes.geometric, angles, quotes}
\usepackage{braket} 		
\usepackage{bm} 			
\usepackage{tensor} 		
\usepackage{slashed} 		
\usepackage{siunitx} 		
\usepackage{lastpage} 		
\usepackage{cite} 			
\usepackage[normalem]{ulem} 
\usepackage{fontawesome} 	
\usepackage{tocloft} 		
\usepackage{titlesec} 		
\usepackage{doi} 			
\usepackage{hyperref} 		
\usepackage[most]{tcolorbox} 					
\usepackage[nameinlink, capitalize]{cleveref} 	
\usepackage[nottoc, notlot, notlof]{tocbibind} 	
\usepackage[ruled, vlined]{algorithm2e} 		
\usepackage{makecell}
\usepackage{wrapfig}

\hypersetup{
	pdftitle={Lorentz-Equivariance without Limitations},
	pdfauthor={Luigi Favaro, Gerrit Gerhartz, Fred A. Hamprecht, Peter Lippmann, Sebastian Pitz, Tilman Plehn, Huilin Qu, Jonas Spinner},
	colorlinks=true, 			
	linkcolor={red!50!black}, 	
	citecolor={blue!50!black}, 	
	urlcolor={blue!80!black} 	
} 

\DeclareSymbolFont{usualmathcal}{OMS}{cmsy}{m}{n}
\DeclareSymbolFontAlphabet{\mathcal}{usualmathcal}

\newlist{todolist}{itemize}{2}
\setlist[todolist]{label=$\square$}
\usepackage{pifont}
\SetArgSty{textnormal}
\SetKwComment{Comment}{{\small\#}~}{}
\SetCommentSty{mycommfont}

\setitemize{itemsep=0pt, parsep=0pt} 				
\setenumerate{itemsep=0pt, parsep=0pt} 				
\setitemize{itemsep=2pt,topsep=2pt,parsep=0pt,partopsep=0pt,leftmargin=*}
\setenumerate{itemsep=0pt,topsep=2pt,parsep=0pt,partopsep=0pt,labelindent=3pt,leftmargin=*}

%% file: incl_shortcuts.tex
\newcommand{\pl}{p_\text{latent}}
\newcommand{\pd}{p_\text{data}}

\newcommand{\loss}{\mathcal{L}}

\newcommand{\Langle}{\big\langle}
\newcommand{\Rangle}{\big\rangle}

\newcommand\one{\leavevmode\hbox{\small1\normalsize\kern-.33em1}}
\newcommand{\p}{\partial}

\newcommand{\qqquad}{\qquad \qquad}

\newcommand{\gev}{\text{GeV}}

\def\slashchar#1{\setbox0=\hbox{$#1$}           
   \dimen0=\wd0                                 
   \setbox1=\hbox{/} \dimen1=\wd1               
   \ifdim\dimen0>\dimen1                        
      \rlap{\hbox to \dimen0{\hfil/\hfil}}      
      #1                                        
   \else                                        
      \rlap{\hbox to \dimen1{\hfil$#1$\hfil}}   
      /                                         
   \fi}

\newcommand{\dz}{\phantom{0}}
\newcommand{\result}[2]{{#1}~\textcolor{error}{$\pm$ {#2}}}

\definecolor{error}{rgb}{0.7, 0.7, 0.7}
\newcommand{\error}[1]{\textcolor{error}{{#1}}}

\usepackage{scalerel}
\newcommand{\smboxop}{%
  \mathop{\scaleobj{1.45}{\square}}\displaylimits}